\documentclass[apJ]{emulateapj}
\usepackage{textcomp}
\usepackage{epsfig, epstopdf}
\usepackage{amssymb,amsmath}
\usepackage{longtable,lscape}
\usepackage[]{natbib}
\newcommand{\lRL}{\hbox{log~$R_L$}}

\newcommand{\lx}{\hbox{$L_{\rm X}$}}
\newcommand{\lLa}{\hbox{log~$L_{2-10}$}}

\newcommand{\kms}{\hbox{km~s$^{-1}$}}
\newcommand{\cmsq}{\hbox{cm$^{-2}$}}

\newcommand{\flux}{\hbox{erg~cm$^{-2}$~s$^{-1}$ }}
\newcommand{\lumin}{\hbox{erg~s$^{-1}$}}

\newcommand{\aox}{\hbox{$\alpha_{\rm ox}$}}
\newcommand{\nh}{\hbox{${N}_{\rm H}$}}

\newcommand{\RF}{\hbox{rest-frame}}

\newcommand{\Cstat}{\hbox{$C$-statistic}}
\newcommand{\XR}{\hbox{X-ray}}
\newcommand{\PL}{\hbox{power-law}}
\newcommand{\chandra}{\emph{Chandra}}

\newcommand{\xmm}{\hbox{\emph{XMM-Newton}}}

\newcommand{\xspec}{{\sc xspec}}

\newcommand{\marx}{{\sc marx}}

\newcommand{\chasa}{PMN~J0235$-$1805}
\newcommand{\chasb}{PMN~J2219$-$2719}

\newcommand{\xmmsa}{SDSS~J0011$+$1446}
\newcommand{\xmmsb}{PMN~J0214$-$0518}
\newcommand{\xmmsc}{SDSS~J0839$+$5112}
\newcommand{\xmmsd}{SDSS~J1309$+$5733}

\begin{document}

\title{THE X-RAY PROPERTIES OF TYPICAL HIGH-REDSHIFT RADIO-LOUD QUASARS}

\author{C.~Saez\altaffilmark{1}, W.~N.~Brandt\altaffilmark{1}, O.~Shemmer\altaffilmark{2},  L.~Chomiuk\altaffilmark{3,4}, L.~A.~Lopez\altaffilmark{5}, H.~L. Marshall\altaffilmark{6}, B.~P.~Miller\altaffilmark{7} \& C.~Vignali\altaffilmark{8,9}}

\altaffiltext{1}{Department of Astronomy \& Astrophysics, Pennsylvania State University,
University Park, PA 16802, saez@astro.psu.edu}
\altaffiltext{2}{Department of Physics, University of North Texas, Denton, TX 76203, USA}

\altaffiltext{3}{Jansky Fellow of the National Radio Astronomy Observatory, Charlottesville, VA 22903, USA}

\altaffiltext{4}{Harvard-Smithsonian Center for Astrophysics, 60 Garden Street, Cambridge, MA 02138}

\altaffiltext{5}{Department of Astronomy and Astrophysics, University of California Santa Cruz, 159 Interdisciplinary Sciences Building, 1156 High Street, Santa Cruz, CA 95064, USA}
\altaffiltext{6}{Kavli Institute for Astrophysics and Space Research, Massachusetts Institute of Technology, 77 Massachusetts Ave., Cambridge, MA 02139, USA}
\altaffiltext{7}{Department of Astronomy, University of Michigan, 500 Church Street, Ann Arbor, MI 48109, USA}

\altaffiltext{8}{Dipartimento di Astronomia, Universit\`a degli Studi di Bologna, Via Ranzani 1, 40127 Bologna, Italy}
\altaffiltext{9}{INAF Ð Osservatorio Astronomico di Bologna, Via Ranzani 1, 40127 Bologna, Italy}
\begin{abstract}

 We report spectral, imaging, and variability results from four new 
\xmm\ observations and two new \chandra\ observations of 
high-redshift ($z \gtrsim 4$) radio-loud quasars (RLQs). Our targets 
span lower, and more representative, values of radio loudness than 
those of past samples of high-redshift RLQs studied in the \XR\ 
regime. Our spectral analyses show \PL\ \XR\ continua with 
a mean photon index, $\langle \Gamma \rangle=1.74\pm 0.11$, that is consistent with 
measurements of lower redshift RLQs. These continua are likely 
dominated by jet-linked \XR\  emission, and they follow the expected
anti-correlation between photon index and radio loudness. We find 
no evidence of iron~K$\alpha$ emission lines or Compton-reflection continua. 
Our data also constrain intrinsic \XR\ absorption in these RLQs. 
We find evidence for significant absorption 
($N_{\rm H}\approx 1.7\times 10^{22}~\cmsq$) in one RLQ of our 
sample (\xmmsa); the incidence of \XR\ absorption in our sample appears plausibly
consistent with that for high-redshift RLQs that have higher values 
of radio loudness. In the \chandra\ observation of \chasb\ we detect  apparent extended  ($\sim 14$~kpc) \XR\ emission  that is most likely due to 
a jet; the \XR\ luminosity of this putative jet is $\approx 2$\% 
that of the core.  The analysis of a 4.9~GHz VLA image of \chasb\ reveals a structure that matches the \XR\ extension found in this source. We also find evidence for long-term 
\hbox{(450--460~days)} \XR\ variability by \hbox{80--100\%} in two of 
our targets.

\end{abstract}


\section{INTRODUCTION}


Active Galactic Nuclei (AGN) are tracers of supermassive black hole (SMBH) growth in the centers of galaxies. This activity can be used to find and probe the evolution of massive galaxies at cosmological epochs close to the period of reionization \citep[e.g.,][]{2006ARA&A..44..415F}.  The study of high-redshift quasars ($z \gtrsim 4$)  is particularly important because it provides information on the period when the first SMBHs were formed in the Universe. \XR\ studies reveal conditions in the immediate 
vicinity of these SMBHs as well as their larger scale environment (e.g., with intrinsic or intervening absorption).  Most \XR\ studies of high-redshift quasars have focused on radio-quiet quasars (RQQs; those with radio-loudness\footnote{The radio-loudness parameter ($R_L$) is defined as the ratio between the flux density  at 5~GHz and the flux density  at 4400 \AA\  in the rest frame of the source \citep{1989AJ.....98.1195K}.} parameter $R_L \lesssim 10$). The main result of these studies is that high-redshift RQQs have similar \XR\ properties (e.g., \XR\ \PL\ photon index, $\Gamma$, and optical-to-\XR\  \PL\ slope\footnote{The optical-to-\XR\ \PL\ slope is defined as $\aox= {\rm log}(f_{\rm 2keV}/f_{\hbox{\tiny 2500\AA}})/{\rm log}(\nu_{\rm 2keV}/\nu_{\hbox{\tiny 2500\AA}}) $ \citep{1979ApJ...234L...9T}.}, \aox) to those at lower redshift ($z=1-3$) of matched luminosity  \citep[e.g.,][]{2006ApJ...644...86S, 2007ApJ...665.1004J}.\footnote{For RQQs, \aox\ shows a dependence upon the UV luminosity  \citep[e.g.,][]{2006AJ....131.2826S}. For these sources there is also evidence that $\Gamma$ may  depend on the \XR\ luminosity \citep[e.g.,][]{2008AJ....135.1505S}  and/or the accretion rate  \citep[in terms of the Eddington ratio;  e.g.,][]{2008ApJ...682...81S}.}  Thus, high-redshift RQQs seem to be feeding and growing in 
 a generally similar manner to those at lower redshift. 


Radio-loud quasars (RLQs; those with $R_L \gtrsim 10-100$) have strong radio emission which is usually linked to the presence of jets emitted from the AGN at relativistic speeds \citep[e.g.,][]{1984RvMP...56..255B}. \XR\ spectral and variability studies of high-redshift RLQs provide a probe of their pc-scale jet-linked \XR\ emission, since this jet-linked emission often dominates the observed \XR\ luminosity.  
Due to this jet-linked component, RLQs have stronger \XR\ emission than RQQs of similar optical/UV luminosities \citep[e.g.,][]{1981ApJ...245..357Z}. This excess in \XR\ emission increases with the radio loudness \citep[e.g.,][]{1987ApJ...313..596W}; for example, for ${\rm log}~R_L \sim 4$ the \XR\ luminosity of RLQs can be up to $\sim10$ times higher than the luminosity of RQQs with similar optical/UV luminosity \citep[e.g.,][]{2011ApJ...726...20M}.  The jet-linked emission of RLQs also makes their \XR\ spectra harder as $R_L$ increases \citep[e.g.,][]{1987ApJ...323..243W, 2000MNRAS.316..234R} .
RLQs represent  \hbox{10--20\%} of the whole quasar population \citep[e.g.,][]{2002AJ....124.2364I, 2007ApJ...656..680J}. Notably, the frequency of radio jets in luminous quasars seems to 
 change significantly with redshift and UV luminosity  \citep[e.g.,][]{ 2007ApJ...656..680J}.
 Thus, it is of interest to assess if the physical processes operating
 within jets also change with redshift; any such changes should probably be most apparent at the highest redshifts.  RLQs at $z\gtrsim 4$ have been studied much less well in the
 \XR\ regime than correspondingly high-redshift RQQs. The number of sensitive exploratory \XR\ observations (and thus
  \XR\ detections) is much smaller ($20-30$ in total), and the number of RLQs with 
quality \XR\ spectra is even smaller still ($\lesssim 10$).  Those $z \gtrsim 4$ RLQs that have been targeted for \XR\ 
 spectroscopy are usually highly radio loud \citep[$R_L > 1000$; e.g.,][]{2004MNRAS.350..207W, 2006MNRAS.368..985Y}. Such highly 
 radio-loud objects make up only a small fraction ($\lesssim 5$\%) of the RLQ
 population, so they are not representative objects. 

X-ray observations of high-redshift RLQs with good angular resolution can also facilitate study of kpc-scale \XR\ jets.
  Two of the most popular models for this kpc-scale emission \citep[e.g.,][and references therein]{2009A&ARv..17....1W} are (1) an extra high-energy
synchrotron component, and (2) inverse Compton (IC) scattering by relativistic electrons of photons of the cosmic
microwave background (CMB; or the  IC/CMB model). 
 For synchrotron models, we do not obviously  expect that the \XR\ emission properties would depend on redshift.
The energy density of the CMB increases as $(1+z)^4$; as a consequence, an implication of the IC/CMB model is that for  high-redshift RLQs  ($z \gtrsim 4$) the \XR\ emission from extended jets may outshine that from the core  \citep[e.g.,][]{2002ApJ...569L..23S}. High-redshift sources with such dominant \XR\ emission from extended jets have not yet been reported \citep[e.g.,][]{2006AJ....131.1914L}. This IC/CMB enhancement of the \XR\ emission could also be observable via noticeable systematic changes with redshift of \aox\ and the radio-to-\XR\  \PL\ slope\footnote{The radio-to-\XR\ \PL\ slope is defined as $\alpha_{rx}= {\rm log}(f_{\rm 1keV}/f_r)/{\rm log}(\nu_{\rm 1keV}/\nu_r) $; in \cite{2011ApJS..193...15M} the radio frequency used is $\nu_r=8.64$~GHz.}, $\alpha_{rx}$; such increments have not been observed \citep[e.g.,][]{2005ApJS..156...13M, 2011ApJS..193...15M, 2011ApJ...726...20M}. 

The potential \XR\ absorption in  representative high-redshift RLQs is also a critical topic to study.
Changes in the strength and frequency of RLQ
intrinsic \XR\ absorption (i.e., $N_{\rm H}$) with redshift have been discussed 
 by many authors  \citep[e.g.,][]{1994ApJ...422...60E, 1997ApJ...478..492C,1998ApJ...492...79F, 2000MNRAS.316..234R,  2005MNRAS.364..195P,  2006MNRAS.368..985Y}.  These studies demonstrate that the fraction of RLQs showing low-energy \XR\ cutoffs appears to 
 rise with redshift. These cutoffs are typically attributed to the
 presence of intrinsic \XR\ absorption. Column densities of $N_{\rm H}\approx$~(1--3)$\times 10^{22}$~cm$^{-2}$ 
 are often seen at $z\gtrsim 2$.  RQQs do not show clear evidence for increasing \XR\ absorption 
 with redshift; this contrast with RLQs demonstrates that there must be a connection between \XR\ absorption and radio loudness. 
The absorbing gas is thought to be associated with the 
 environments of RLQs, but its precise nature is unclear: it may be 
 circumnuclear, located in the young host galaxy, or entrained by the 
 radio jets.  Based on Swift/GRB data set it has also been proposed recently that this absorption may arise in the  diffuse
intergalactic medium  \citep[IGM; e.g.,][]{2011ApJ...734...26B}. Three of the seven highly \hbox{radio-loud} ($\lRL \gtrsim 3$) quasars at $z>4$ with \XR\ spectra 
 show evidence for substantial \XR\ absorption  \citep[e.g.,][]{2001MNRAS.323..373F,  2004MNRAS.350..207W, 2005MNRAS.364..195P, 2006MNRAS.368..985Y}, a finding 
 that is broadly consistent with an extrapolation of the trend observed 
 at lower redshifts. However, the source statistics are limited, 
 and it is precarious to generalize the results from a few extreme highly radio-loud
 objects to the high-redshift RLQ population as a whole.  Note also that moderate \hbox{low-energy}
absorption could be mimicked by a curved
intrinsic spectrum; this possibility cannot be ruled out for RLQs \citep[e.g.,][]{2007ApJ...665..980T}.


In past work, we have targeted representative flat-spectrum RLQs at $z \approx 3.5-5$, typically having moderate $R_L$ values ($R_L \sim 90-400$), using \chandra\ snapshot observations \citep[5--10 ks;][]{2004AJ....128..523B, 2006AJ....131.1914L}.
 These snapshot observations gave information about the level of \XR\
 emission and allowed basic population studies. However, they did not allow 
 detailed physical studies of individual objects. 
Thus, in order to investigate some of the issues described above, we 
 have targeted selected objects from our snapshot samples for longer
 follow-up observations with \xmm\ and \chandra. These observations
 provide spectral constraints as well as information about long-term \XR\
 variability. Furthermore, the observations with \chandra\ allow effective
 searches for kpc-scale \XR\ jet emission. 
 We were awarded time to observe four targets with \xmm\ and two 
 targets with \chandra. The two targets observed with \chandra\ were 
selected to follow-up hints of \XR\ extension based upon analyses of the
 snapshot data.  We supplemented the follow-up 
\chandra\ imaging with high-resolution Very Large Array observations to 
constrain the presence and extent of putative jets.
 
The layout of this paper is as follows: in \S 2 we describe our sample, the \XR\ data reduction, and the radio data reduction;  in \S 3 we provide a discussion of the most important results found in this work; and in \S 4 we summarize our results and main conclusions.  Throughout this paper, unless stated
otherwise,  we use cgs units, the errors
listed are at the 1$\sigma$ level, and we adopt a flat
$\Lambda$-dominated universe with $H_0=70~\kms$~Mpc$^{-1}$,
$\Omega_\Lambda=0.7$, and $\Omega_M=0.3$.

\begin{deluxetable*}{ccccccccc}
\tablecolumns{9}
\tabletypesize{\scriptsize}
 \tablewidth{0pt}
\tablecaption{Log of \xmm\ Observations \label{tab:xmmo}}
\tablehead
{
\colhead{} &
\colhead{} &
\colhead{} &
\colhead{} &
\multicolumn{3}{c}{\underline{{\sc exposure time\tablenotemark{b}} (ks) / {\sc total counts\tablenotemark{c}}}} \\
\colhead{\sc object name} &
\colhead{\sc $\alpha_{2000.0}$\tablenotemark{a}} &
\colhead{\sc $\delta_{2000.0}$\tablenotemark{a} }&
\colhead{\sc obs. date / ID} &
\colhead{MOS1} &
\colhead{MOS2} &
\colhead{pn} 
}
\startdata
\xmmsa  &  00 11 15.23 & $+$14 46 01.8  &
2010Jan07/0600090101 & 29.3(38.5)/$761_{-28}^{+29}$  & 29.3(38.5)/$773_{-28}^{+29}$ & 24.5(36.4)/$2336_{-48}^{+49}$  \\
\xmmsb    & 02 14 29.30 & $-$05 17 44.6 &
2010Jan01/0600090401 & 58.9(60.5)/$186_{-14}^{+15}$ & 54.3(60.5)/$205_{-14}^{+15}$ & 46.0(58.9)/$773_{-28}^{+29}$  \\
\xmmsc  & 08 39 46.21 & $+$51 12 02.8  &
2006Apr12/0301340101& 9.6(26.6)/$101_{-10}^{+11}$  & 9.6(26.6)/$87_{-9}^{+10}$ & $7.4(24.9)/314_{-18}^{+19}$  \\
\xmmsd   & 13 09 40.69 & $+$57 33 09.9  &
2006Apr22/0301340501 & 29.4(54.6)/$85_{-9}^{+10}$ & 28.6(54.6)/$112_{-11}^{+12}$ & 18.3(52.9)/$235_{-15}^{+16}$  \\

\enddata

\tablecomments{Errors on the X-ray counts were computed according to Tables 1 and 2 of \cite{1986ApJ...303..336G}.}

\tablenotetext{a}{Optical positions in J2000.0 equatorial coordinates. The  positions of \xmmsa\ and \xmmsc\ and \xmmsd\ are presented in \cite{2007ApJS..172..634A}, and the position of \xmmsb\ is presented in \cite{2002ApJS..143....1M}.  }

\tablenotetext{b}{The exposure times and photon counts are obtained after screening the data for flaring. The exposure times before the screening are in parentheses.}

\tablenotetext{c}{Total source counts in a circular region of radius of 30$\arcsec$ centered on the X-ray source.  For the MOS cameras and pn camera the photon energies in consideration are the 0.5--10~keV  and the 0.3--10~keV observed band respectively.}

\end{deluxetable*}

\begin{deluxetable*}{ccccccc}
\tablecolumns{7}
\tabletypesize{\scriptsize}
 \tablewidth{0pt}
\tablecaption{Log of \chandra\ Observations \label{tab:chao}}
\tablehead
{
\colhead{} & \colhead{} & \colhead{} & \colhead{} & \colhead{{\sc exposure time}\tablenotemark{b}}  & \\
\colhead{\sc object name} & \colhead{$\alpha_{2000.0}$\tablenotemark{a}} & \colhead{$\delta_{2000.0}$\tablenotemark{a}} & \colhead{\sc observation date/id} & \colhead{(ks)} & \colhead{\sc total counts\tablenotemark{b,c}~~}  & \colhead{Ref.\tablenotemark{d}}
}
\startdata

\multicolumn{7}{c}{\small New observations} \\
 \\
\hline
\chasa  & 02 35 02.50 & $-$18 05 51.0  &
2009 Feb 22 / 10306  & 19.9(21.5) &  141$_{-12}^{+13}$ & 1\\
\chasb & 22 19 35.32 & $-$27 19 03.3  &
2009 Sep 12 / 10305 & 37.9(38.5) & 1803$_{-42}^{+43}$  & 1 \\

\hline
\\
\multicolumn{7}{c}{\small Archival observations} \\
\\
\hline
\xmmsa  &  00 11 15.23 & $+$14 46 01.8& 2003 May 20 / 3957 & 3.5(3.8) & 135$_{-12}^{+13}$ & 2\\
\xmmsb    & 02 14 29.30 & $-$05 17 44.6 & 2003 Nov 26 / 4767 & 4.0(4.3) & 44 $_{-7}^{+8}$ & 3\\
\chasa    &  02 35 02.50 & $-$18 05 51.0  &
2004 Feb 07 / 4766 & 3.9(3.9) &  20$\pm$5 & 3 \\
\xmmsc  & 08 39 46.21 & $+$51 12 02.8 & 2003 Jan 23 / 3562 & 4.9(5.0) & 82$_{-9}^{+10}$  & 4\\
\xmmsd   & 13 09 40.69 & $+$57 33 09.9 & 2003 Jul 12 / 3564 & 4.6(4.7) & 31$_{-6}^{+7}$ & 4 \\
\chasb   &  22 19 35.32 & $-$27 19 03.3  &
2003 Nov 19 / 4769 & 8.1(8.2) & 226$_{-15}^{+16}$ & 3 \\

\enddata
\tablecomments{Errors on the X-ray counts were computed according to Tables 1 and 2 of \cite{1986ApJ...303..336G}.}
\tablenotetext{a}{Optical positions in J2000.0 equatorial coordinates. The positions of \chasa\ and \chasb\ are  presented in \cite{2002ApJS..143....1M}.}
\tablenotetext{b}{ The exposure times and photon counts are obtained after the screening of the data. The exposure times before the screening are in parenthesis.}
\tablenotetext{c}{ Total source counts with energies in the 0.5--8~keV observed band  in a circular region of radius of 3.9 arcsec (8 pixels) centered on the X-ray source. }
\tablenotetext{d}{{\sc References:} (1) This work; (2) \cite{2006ApJ...644...86S}; (3) \cite{2006AJ....131.1914L}; (4) \cite{2004AJ....128..523B}.}

\end{deluxetable*}

\begin{deluxetable*}{ccccccccccc}
\tablecolumns{9}
\tabletypesize{\scriptsize}
 \tablewidth{0pt}
\tablecaption{Optical, X-ray and Radio Properties of Surveyed Radio-Loud Quasars at High Redshift \label{tab:opra}}
\tablehead
{
\colhead{} & \colhead{} & \colhead{} & \colhead{$S_{1.4 \rm GHz }$\tablenotemark{b}} & \colhead{} & \colhead{}  &  \colhead{}  \\
\colhead{\sc object name} & \colhead{$z$} & \colhead{AB$_{1450}$\tablenotemark{a}} & \colhead{(mJy)} & \colhead{$\alpha_r$\tablenotemark{c}}  &  \colhead{\aox \tablenotemark{d}} & \colhead{log~$R_L$\tablenotemark{e}} &  \colhead{log~$L_{\rm 5 GHz}$\tablenotemark{f}} & \colhead{log~$L_{\rm 2500}$\tablenotemark{g}} & \colhead{log~$L_{\rm 2  keV}$\tablenotemark{h}} & \colhead{$\Delta$log~$L_{\rm 2  keV}$\tablenotemark{i}} 
}
\startdata
\xmmsa & 4.97  & 18.1   & 36 &  $-$0.12 &  $-$1.31$\pm$0.03 & 2.01 & 34.21  & 32.08  & 28.66  & 1.10\\
\xmmsb & 3.99 & 18.8  &   41 &  $+$0.65  &$-$1.53$\pm$0.04  & 2.25 & 34.01 & 31.65 & 27.67 & 0.41 \\
\chasa &  4.31 & 18.9 &  36  & $+$0.26 & $-$1.50$\pm$0.04  & 2.28 & 34.05 & 31.64 & 27.74 & 0.49  \\
\xmmsc & 4.39 & 18.8 & 43  & $+$0.14 &  $-$1.40$\pm$0.03 & 2.33 & 34.16 & 31.71 & 28.07  &  0.77\\
\xmmsd & 4.27 &  19.3 &  11 & ... & $-$1.51$\pm$0.03  &1.98 &   33.58   & 31.48  & 27.53  & 0.39 \\
\chasb & 3.63 &    20.3 &  304 & $-$0.26 & $-$1.09$\pm$0.03  & 3.85 &   34.93  &   30.96 &    28.11 &     1.34  \\


\enddata

\tablenotetext{a}{The monochromatic $AB$ magnitude at rest-frame wavelength 1450 \AA. For \xmmsb, \chasa, and \chasb\ we used the $R$ magnitudes found in  \cite{2002A&A...391..509H} to calculate $AB_{1450}$ by using the  empirical relation  $AB_{1450} = R - 0.684z + 3.10$.  For \xmmsa,  \xmmsc, and \xmmsd\ we used the $i$-band SDSS magnitudes to calculate $AB_{1450}$  by using the  empirical relation  $AB_{1450}= i - 0.2 $. Notice that these estimates of $AB_{1450}$ are reliable within $\approx$~0.1--0.2 mag for the redshifts of our sources.}
\tablenotetext{b}{The 1.4~GHz flux density;  from the NRAO VLA Sky Survey (NVSS) \citep{1998AJ....115.1693C}. For sources covered by VLA FIRST Sky survey \citep{1997ApJ...475..479W}, their 1.4~GHz flux density  differ in less than 10\% with the values presented.}
\tablenotetext{c}{The radio power-law slope between 1.4 and 5 GHz (observed frame). The 5~GHz flux density is from the Green Bank 6 cm \citep[GB6;][]{1996ApJS..103..427G} or Parkes-MIT-NRAO \citep[PMN;][] {1993AJ....105.1666G} surveys. }
\tablenotetext{d}{The optical-to-X-ray power-law slope $\aox= {\rm log}(f_{\rm 2keV}/f_{\hbox{\tiny 2500\AA}})/{\rm log}(\nu_{\rm 2keV}/\nu_{\hbox{\tiny 2500\AA}}) $. The average difference between  measured and  predicted \aox\ for RQQ is $\langle \Delta \aox \rangle \approx 0.3$, based on \cite{2006AJ....131.2826S} \aox-$L_{\hbox{\tiny 2500\AA}}$ relation.}

\tablenotetext{e}{Radio-loudness parameter, defined as $R_L= f_{5GHz}/f_{\hbox{\tiny 4400\AA}}$ \citep[rest frame;][]{1989AJ.....98.1195K}. The flux densities at 5~GHz in the rest-frame were obtained from the flux densities at 1.4~GHz in the observed-frame assuming a \PL\ spectrum with spectral slope given by $\alpha_r$; when $\alpha_r$ is not available we assume $\alpha_r=0$. The fluxes at 4400 \AA\ in the rest-frame are obtained from the AB magnitudes assuming a \PL\ spectrum with spectral slope of  $-$0.5  \citep{2001AJ....122..549V}. }
\tablenotetext{f}{\hbox{Logarithm} of the monochromatic luminosity at  5~GHz. The luminosities were obtained from the flux densities at 1.4~GHz in the observed-frame assuming a \PL\ spectrum with spectral slope given by $\alpha_r$; when $\alpha_r$ is not available we assume $\alpha_r=0$. }
\tablenotetext{g}{\hbox{Logarithm} of the monochromatic luminosity at 2500 \AA. These were computed from the $AB_{1450}$ assuming and optical \PL\ of $\alpha=-0.5$ \citep{2001AJ....122..549V}.} 
\tablenotetext{h}{\hbox{Logarithm} of the absorption-corrected monochromatic luminosity at 2~keV.} 

\tablenotetext{i}{The difference between measured and the predicted log~$L_{\rm 2  keV}$ for RQQs;  based on the \cite{2007ApJ...665.1004J} relation \hbox{${\rm log}~L_{\rm 2  keV}=0.709 \times {\rm log}~L_{\rm 2500}+4.822$} .}

\end{deluxetable*}

\begin{figure}
   \includegraphics[width=8.6cm,height=7.8cm]{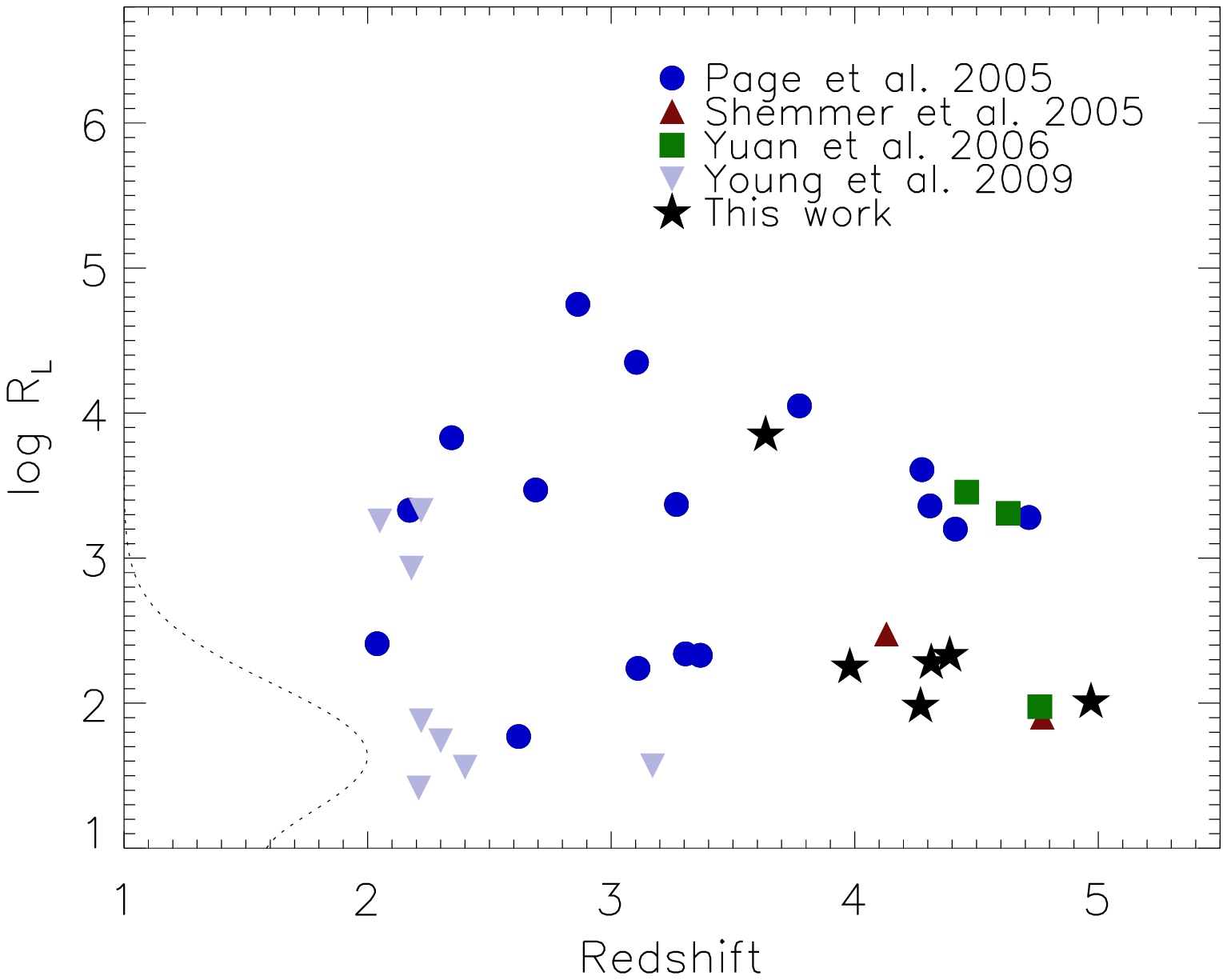}
        \centering
       \caption{Radio-loudness parameter versus redshift for our sample (stars) and $z>2$ RLQs observed with \chandra\ and \xmm\ (circles are from \citeauthor{2005MNRAS.364..195P}~2005, triangles from \citeauthor{2005ApJ...630..729S}~2005, squares from \citeauthor{2006MNRAS.368..985Y}~2006, and inverted triangles from  \citeauthor{2009ApJS..183...17Y}~2009). The dotted curve along the left side of the plot shows the relative number of RLQs versus $R_L$ from \cite{2002AJ....124.2364I}. Note that our sample significantly improves the source statistics for moderately radio-loud quasars at the highest redshifts.  }
     \label{fig:zRLs}
     \end{figure}

\section{OBSERVATIONS AND DATA REDUCTION}

\subsection{X-ray data}

In this work we perform analyses of four new \xmm\ observations and two new  \chandra\ observations of high-redshift ($z = 3.6-5.0$) RLQs with moderate $R_L$. These observations were obtained as a follow-up of \chandra\  snapshot observations \citep{ 2004AJ....128..523B, 2006AJ....131.1914L, 2006ApJ...644...86S}. The original snapshot observations were also reprocessed with the main objective of studying any long-term \XR\ variability.
 The observation logs of the sources analyzed in this work, that include
observation dates, observed total counts, total exposure times,
and observational identification numbers, are presented in Table~\ref{tab:xmmo} for the \xmm\ observations and Table~\ref{tab:chao} for the \chandra\ observations.  Additionally, in Table~\ref{tab:opra} we  present basic optical, \XR, and radio properties for our sample;  among these are the redshift and $R_L$ of each source. 
The radio-loudness parameters range from $\approx 90-7000$ with five out of  the six sources having $R_L \lesssim 200$. 
From Table~\ref{tab:opra} we also notice that our RLQ sample is on average $\sim  6$ times brighter in \hbox{X-rays} than RQQs of similar UV luminosity  (based on \citeauthor{2007ApJ...665.1004J} 2007 and \citeauthor{2011ApJ...726...20M} 2011).  This is suggestive evidence that the \XR\ continua of our sample are dominated by jet-linked emission.
In Figure~\ref{fig:zRLs} we have plotted our sources (with stars) in a diagram of $\lRL$ vs. $z$. We also have added to this plot RLQs from other high-redshift samples \citep{2005MNRAS.364..195P, 2005ApJ...630..729S, 2006MNRAS.368..985Y, 2009ApJS..183...17Y} with high-quality \XR\ spectra.\footnote{In the added sample of RLQs there are some sources that have been analyzed in more than one work. The sources with overlap are PMN~J0525$-$3343, RX~J1028.6$-$0844, SDSS~J1430$+$4204  \citep[these three sources were analyzed in][]{2005MNRAS.364..195P, 2006MNRAS.368..985Y} and SDSS~J1510$+$5702 \citep[analyzed in][]{2005MNRAS.364..195P, 2006MNRAS.368..985Y, 2009ApJS..183...17Y}. We avoid repeated sources in our extended sample by selecting firstly the spectral properties of the sources presented in \cite{2005MNRAS.364..195P}, secondly those presented in \cite{2006MNRAS.368..985Y}, and thirdly those presented in \cite{2009ApJS..183...17Y}.  }  
Our sample  is concentrated toward high redshifts and typically representative  radio-loudness parameters, where it significantly improves the source statistics. This can be seen from Figure~\ref{fig:zRLs}  by comparing the $R_L$ of our sources to the  $R_L$ distribution found from SDSS/FIRST detections  by  \cite{2002AJ....124.2364I}   (dotted-line curve).

The \xmm\ data were analyzed with the Science Analysis Software (SAS) version 10.0.0 provided
by the \xmm\ Science Operations Centre (SOC).  The event files were obtained by including the events with ${\rm flag }=0$, ${\rm pattern} \leq 12$ (${\rm pattern} \leq 4$), and $200 \leq {\rm PI} \leq 12,000$ ($150 \leq {\rm PI} \leq15,000$) for the MOS (pn) detectors. The event files were also temporally  filtered in order to remove periods of flaring activity.\footnote{To filter the event files of periods of flaring activity, we create light curves using single events with energy greater than 10~keV.  Using these light curves we find and remove from the event files the time periods where the count rate is above $\sim$0.5~count~s$^{-1}$ for the MOS cameras and $\sim$1~count~s$^{-1}$ for the pn camera.}  This flaring activity  tends to saturate the instrumental background and therefore reduces the good-time intervals. The flaring is especially strong in the observations of \xmmsc\ and \xmmsd, where the effective observation time has been reduced by 40\% and 50\%, respectively,  when compared with the original exposure time (see Table~\ref{tab:xmmo}). For the three EPIC detectors (MOS1,  MOS2, and pn), the source and background spectra of each quasar were extracted from a circular region with an aperture radius of 30$^{\prime\prime}$ \footnote{ The 1.5 keV encircled energy at 30$^{\prime\prime}$ from the center of a point  source is between 80-90\% for the EPIC cameras. Note that throughout this work we have applied an aperture correction factor  when we calculate fluxes or luminosities.} and an annular source-free region with a 50$^{\prime\prime}$ inner radius  and 100$^{\prime\prime}$  outer radius, respectively.  The numbers of counts  in the source regions for the \xmm\ cameras are presented in Table~\ref{tab:xmmo}.  These counts are selected at energies of 0.5--10~keV  for the MOS cameras and 0.3--10~keV for the pn camera.
The combined numbers of counts in the EPIC cameras range approximately from 500 to 4000 counts. The response files used to analyze the source spectra were created with the SAS tasks {\sc rmfgen} for the redistribution matrix files (RMFs) and {\sc arfgen} for the ancillary response files (ARFs).

\begin{figure*}
   \includegraphics[width=16cm]{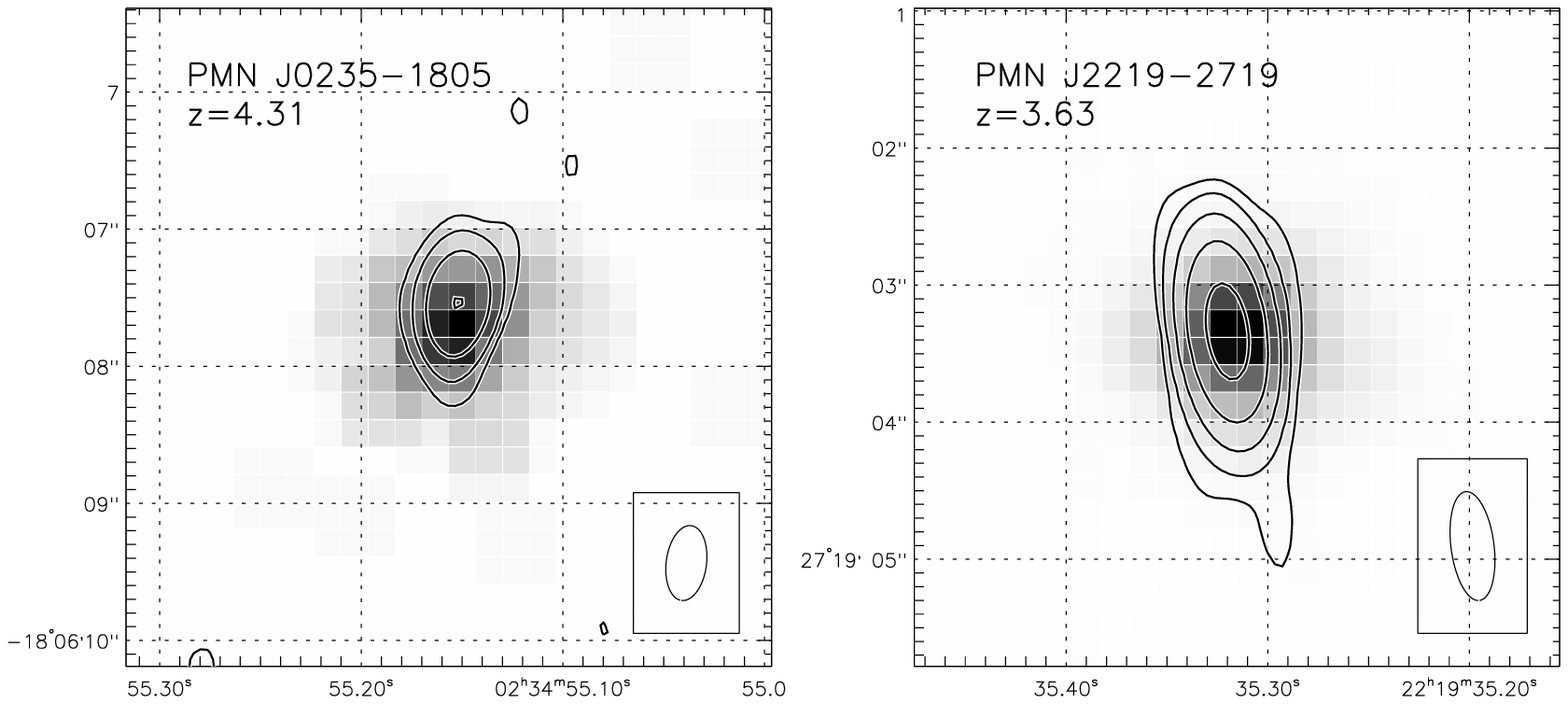}
        \centering
       \caption{Full band (0.5$-$8 keV) \chandra\ images of \chasa\ (left panel) and \chasb\ (right panel). Each panel spans $4.4\arcsec \times 4.4\arcsec$ on the sky; North is up, and East is to the left. Additionally, in each panel the VLA contour plots at 4.9~GHz are shown.
The image resolutions of the radio contour plots are $0.55\arcsec \times 0.30\arcsec$  at ${\rm PA}=-7.7^\circ$ and  $0.80\arcsec \times 0.32\arcsec$ at ${\rm PA}=7.5^\circ$ for \chasa\ and \chasb, respectively. The contour plots begin at 0.4~mJy per beam and increase by factors of 4. Notice that in each panel there is an offset between the \XR\ image and the radio contours of about $0.2\arcsec-0.3\arcsec$; this offset is within the expected astrometric error of  \chandra. The elliptical restoring beams are plotted in the lower-right inset of each panel.}
      \label{fig:char}
     \end{figure*}

The \chandra\ observations (new and archival) were analyzed using the standard software CIAO
4.2 provided by the \chandra\ \XR\ Center (CXC). The telemetry format used was Faint mode for all the observations with the exception of obsID 10306 for which Very Faint mode\footnote{Very Faint mode offers the advantage of reduced background after ground processing.} was used.  Standard CXC threads were employed to screen the data for status, grade (\emph{ASCA} grade 0, 2, 3, 4 and 6 events), and time intervals of acceptable aspect solution and background levels, only. The source and background spectra of each quasar were extracted (using the CIAO task {\sc dmextract}) from a circular region with an aperture radius of 3.9$^{\prime\prime}$~\footnote{The 1.5 keV encircled energy at 3.9$^{\prime\prime}$  from the center of a point  source is $\sim 99\%$ for ACIS-S .}  (8 pixels) and an annular source-free region with an inner radius of 4.9$^{\prime\prime}$ (10 pixels) and an outer radius of  39$^{\prime\prime}$ (80 pixels),  respectively. The numbers of counts  in the source regions at energies of 0.5--8~keV  are presented in Table~\ref{tab:chao}. The \hbox{0.5--8~keV} source counts in the new \chandra\ observations of \chasa\ and \chasb\  are approximately 140 and 1800 counts, respectively (see Table~\ref{tab:chao}). The 0.5--8~keV source counts in the archival observations  range from $\approx 20 - 230$ counts  (see Table~\ref{tab:chao}).  The response files used to analyze the source spectra were created  with the CIAO tasks {\sc mkacisrmf} for the RMFs  and {\sc mkarf} for  the ARFs.

\subsection{Radio data for \chasa\ and \chasb} \label{S:dara}

The new \chandra-observed targets, \chasa\ and \chasb, were observed at the NRAO Very Large Array (VLA) in A configuration, for a nominal resolution of ~1.0$^{\prime\prime}$ and 0.39$^{\prime\prime}$ at 1.4 and 4.9 GHz, respectively. 
\chasa\ was observed in
Aug 2007 (observation proposal code AL696), with 100 minutes spent  at 1.4 GHz and 28 minutes at
4.9 GHz. \chasb\ was observed  on 16 Oct 2004  (observation proposal code AC755), for 35 minutes at 1.4 GHz and 12
minutes at 4.9 GHz. Observations were obtained
in VLA continuum mode, with two intermediate-frequencies (IFs) and full polarization, for a total
bandwith of 86 MHz. 
The data were calibrated using the absolute flux-density calibrators 3C48 and 3C147 and the secondary calibrators 0240$-$231 and
2248$-$325.
Data editing and calibration were carried out in the Astronomical Image Processing System (AIPS). 
The quasars were imaged using uniform weighting for maximal resolution, and were self-calibrated with a phase-only solution interval of
10 sec and amplitude solution intervals of 5 min.
The VLA synthesized beam at 1.4~GHz and 4.9~GHz at the position of \chasa\ has dimensions of 2.04\arcsec$\times$1.03\arcsec\ at ${\rm PA} =-20.2^\circ$  and 0.55\arcsec$\times$0.30\arcsec\ at ${\rm PA}=-7.7^\circ$, respectively. The VLA synthesized beam at 1.4~GHz and 4.9~GHz at the position of \chasb\ has dimensions of 2.70\arcsec$\times$1.00\arcsec\ at ${\rm PA}=3.8^\circ$ and 0.80\arcsec$\times$0.32\arcsec\ at ${\rm PA}=7.5^\circ$, respectively.

\section{RESULTS AND DISCUSSION}

Here we describe and discuss the main results obtained in this work. We start in \S \ref{S:chai} by assessing evidence for extended jets in the \chandra\ images. We continue in \S \ref{S:Span} by providing the main results from our spectral analyses with \xmm\ and \chandra. In \S  \ref{S:vari}  we constrain long and short-term variability. Finally, in \S  \ref{S:corr}  we perform  correlation analyses between several of the parameters that describe our spectral data; in particular we focus on correlations found in low redshift samples, as e.g., $\Gamma$ with $\lRL$ and $\nh$ with z. 

\subsection{Imaging analysis with \chandra \label{S:chai}}

The outstanding angular resolution of \chandra\ ($\sim$~0.5$^{\prime\prime}$)  allows us to study any jet that has an \XR\ spatial extension larger than $\sim 5$~kpc in the RLQ rest frame. 
The analysis of the snapshot observations of  \chasa\  and \chasb\ by  \cite{2006AJ....131.1914L} hinted that these two sources might present extended jets in \hbox{X-rays}.   The putative jets found  by  \cite{2006AJ....131.1914L}  were extended (1$-$2\arcsec) in the western and  south-western direction for \chasa\ and \chasb, respectively. In both sources the \XR\ luminosity of the extended emission was found to be $\approx 3\%$ that of the core \citep{2006AJ....131.1914L} . 

\begin{figure*}
   \includegraphics[width=16cm]{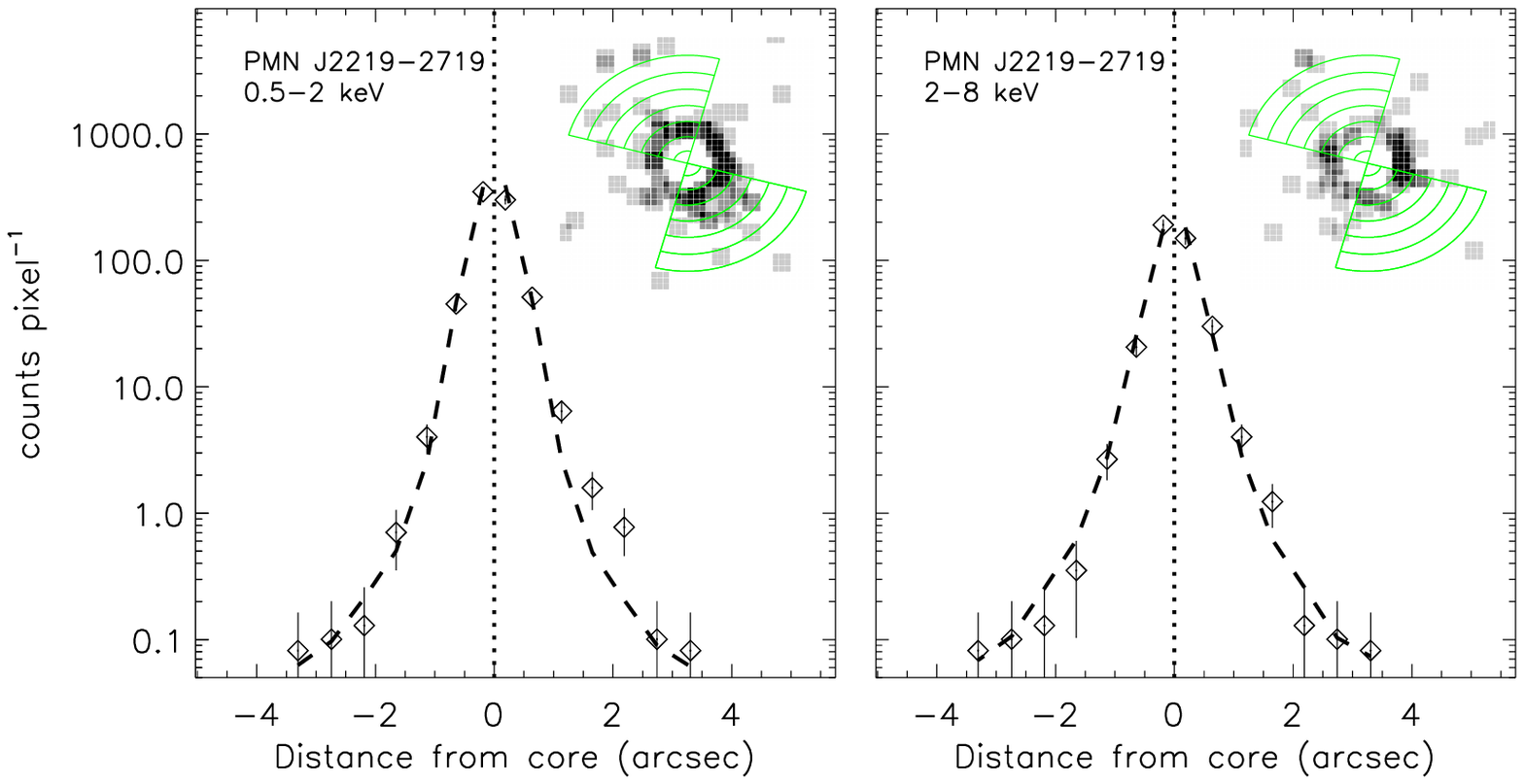}
        \centering
       \caption{0.5$-$2 keV (left panel) and 2$-$8 keV (right panel) radial profiles of the \chasb\ core region, constructed from circular sectors shown in the upper-right inset of each panel. The radial profiles are obtained assuming positive distance in the southwestern direction.  For presentation purposes, in the upper-right inset of each panel,  we have masked the photons coming from a one arc-second radius circular region  centered at the centroid of the source. The dashed curve shows a scaled PSF generated with {\sc marx} with energies close to the average energy of the photons encircled at 2$^{\prime\prime}$ from the centroid of the \XR\ image.}
     \label{fig:cpsf}
     \end{figure*}

In Figure~\ref{fig:char} we present the new full band \XR\ (0.5--8.0 keV) images of \chasa\ and \chasb. In this figure we also have overlaid the contour plots of their 4.9~GHz emission based on the VLA data. From a simple inspection of these images we do not find any obvious evidence for highly extended \XR\ emission in these sources. 
To check further for any evidence of extended emission  we have compared a radial profile extracted
from each of the \chandra\ images using circular annuli  with a point-spread function (PSF) calculated with \marx.\footnote{\marx\  is a \chandra\  ray-tracing simulator (for details see http://space.mit.edu/CXC/MARX) . } 
The images for each source were made using photons in the soft \hbox{0.5--2~keV} band and the hard  \hbox{2--8~keV} band. { These images were obtained by removing pixel randomization and applying the subpixel algorithm of \cite{2004ApJ...610.1204L}.
For each source we obtain a \marx\ simulation assuming a point source with an \XR\ spectrum given by the \hbox{best-fitted} absorbed \PL\ model from the new observations (see \S \ref{S:Span}).}

     \begin{figure}
   \includegraphics[width=8.56cm]{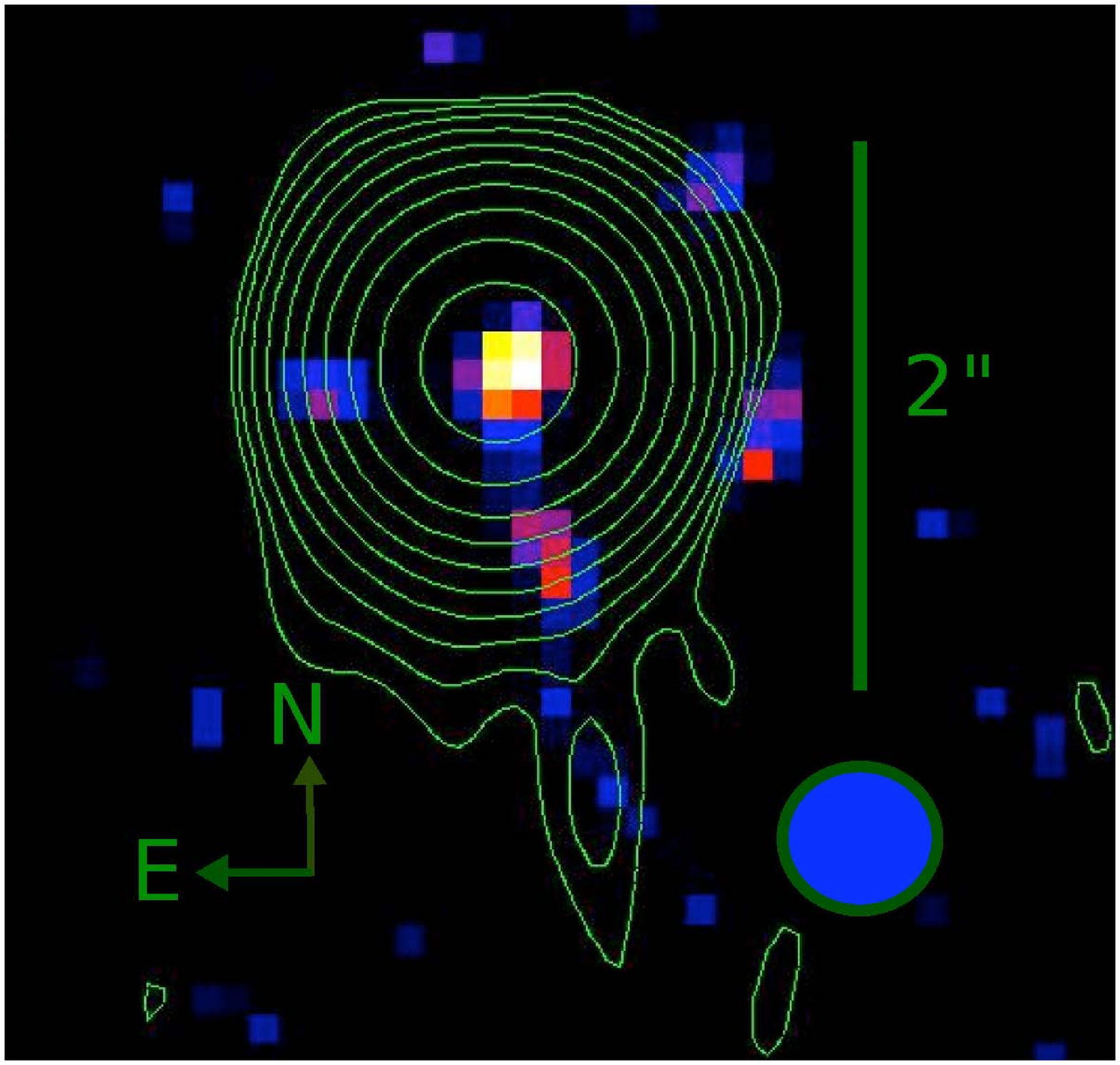}
        \centering
       \caption{Maximum-likelihood reconstruction \citep[see \S2.1 of][]{2006AJ....131.2164T} of the full band \XR\ image of \chasb. The colors yellow-red-blue indicate the decreasing order of photon count densities.  Additionally a super-resolved VLA contour plot at 4.9~GHz is overlaid. 
The image resolution of the radio contour plot is $0.6\arcsec \times 0.6\arcsec$. The restoring beam is the blue circle plotted at the lower-right corner. The contours begin at 0.2 mJy per beam and increase by factors of 2.  }
     \label{fig:rama}
     \end{figure}

From the PSF analysis with \marx\ we found a  hint of  spatial extension in \chasb\ along the south-west direction.  In Figure~\ref{fig:cpsf} the soft band (left panel) and hard band (right panel)  radial profiles of \chasb\ along the south-west direction are presented.  The possible \XR\ spatial extension is seen in the soft band\footnote{The soft band \RF\ coverage for \chasb\ is $\approx$2--9~keV.}  through a count overdensity in the data with respect to the PSF  around 2$^{\prime\prime}$ from the centroid of the source. This count overdensity is seen in two data points with significantly higher counts ($\sim 2\sigma$) than those predicted from the PSF (see left panel of  Figure~\ref{fig:cpsf}).   
 To obtain a basic estimate of the level of putative jet emission in \chasb\ we measure, in the full band (\hbox{0.5--8~keV}), the excess of counts, with respect to a point source, that lie at  $>$1$^{\prime\prime}$ from the core.  For this purpose the radial profile of the source is compared with a \marx\ simulation of a point source. Our results indicate that we have  $38_{-12}^{+13}$ (errors
from Gehrels 1986) additional counts\footnote{For the new \chandra\ observation of \chasb\ the total number of counts in a circular aperture extending from 1\arcsec\ to 4\arcsec\ is $140_{-12}^{+13}$, and the number of counts predicted by our PSF simulation is 102.}  in the data when compared with
the simulated PSF over a circular aperture extending from 1\arcsec\ to
4\arcsec, which implies that the resolved X-ray emission has a
count rate in the full band of $(2.1\pm0.7)\%$ that from the core.
Such a level of putative \XR\ jet emission is consistent with that estimated by Lopez et al. (2006) using the \chandra\ archival observation of \chasb. Additionally, our analyses suggest a south-western \XR\ angular extension of 2\arcsec\ from \chasb\ ($\sim$~14~kpc in the RLQ rest frame), consistent with the possible $\sim$1--2\arcsec\ extension noted by Lopez et al. (2006).
The total number of counts that lie at  $>$1\arcsec\ from the core and that are inside the \hbox{south-western} \hbox{semi-circular} sector presented in the upper-right inset of Figure~\ref{fig:cpsf} is $47_{-7}^{+8}$ \citep[errors from][]{1986ApJ...303..336G}. 
 In the same area the number of expected counts for a point source (from the \marx\ PSF) is $24$. Based on a Poisson distribution with expected mean 24, the null hypothesis of obtaining 47 counts or more is $2.12\times 10^{-5}$.  Therefore the extension found in \chasb\ 
  has a significance of $\approx 4.1 \sigma$.  The putative \XR\ jet  luminosity in the 2--10~keV rest-frame of \chasb\ is $\lLa \sim 44.6$ ($\sim$2\% of the total $L_{2-10}$ luminosity of \chasb; see \S \ref{S:Span}).  
 Given this high \XR\ luminosity, the extended emission for \chasb\ is unlikely to have a non-AGN origin; e.g., starburst activity is typically fainter than $\lLa \sim 42.5$ \citep[see, e.g.,][]{2004AJ....128.2048B}.

\begin{deluxetable*}{cccccccc}
\tablecolumns{8}
\tabletypesize{\scriptsize}
 \tablewidth{0pt}
\tablecaption{Best-fit \hbox{X-ray} spectral parameters \label{tab:xfit}}
\tablehead
{
\colhead{} & \colhead{} & \colhead{\sc galactic \nh\tablenotemark{a}} & \colhead{\nh\tablenotemark{b}} & \colhead{} & \colhead{} &  \\
\colhead{\sc object name} & \colhead{\sc observation id} & \colhead{(10$^{20}$~cm$^{-2}$)} & \colhead{(10$^{22}$~cm$^{-2}$)} & \colhead{$\Gamma$} & \colhead{$f_{0.5-2}$\tablenotemark{c}} & \colhead{\lLa \tablenotemark{d}} & \colhead{$C$-stat(DOF)}
}
\startdata
\multicolumn{8}{c}{\small \xmm\ observations} \\
\\
\hline
 
\xmmsa  & 0600090101 & 4.3 & $1.71^{+0.67}_{-0.64}$  & 1.87$\pm$0.07 &
13.74$\pm$0.75  & 46.62$\pm$0.02   &  $1146.4(1336)$  \\ 
\xmmsb & 0600090401 & 2.6 & $\le1.05$ & 2.10$\pm$0.10 & $1.88 \pm 0.19$   &   45.55$\pm$0.04    & $887.8(903)$  \\
\xmmsc & 0301340101  & 3.3 & $\le 2.33$ & 1.61$\pm$0.12 & $4.42^{+0.73}_{-0.54}$ & 45.99$\pm$0.06     & $407.3(446)$  \\
\xmmsd  & 0301340501 & 1.3 & $\le2.21$  & $1.69^{+0.15}_{-0.17}$ & $1.68\pm0.24$ & 45.53$\pm$0.06   & $478.0(542)$  \\
\hline
\\
\multicolumn{8}{c}{\small \chandra\ observations} \\
\\
\hline
\xmmsa  & 3957 & 4.3 &  $\le 6.1$ & 1.90$\pm$0.26 & $12.31^{+3.98}_{-1.79}$ & 46.58$\pm$0.10 &109.6(90) \\

\xmmsb & 4767 & 2.6 & $\le 4.0$ & $2.16^{+0.53}_{-0.48}$ & $3.72^{+1.23}_{-0.87}$ & 45.85$\pm$0.12 & 24.1(37)  \\

\chasa  &  10306 & 2.7 & $\le4.3$  & 1.86$\pm$0.25 &
$2.32\pm0.37$  & 45.71$\pm$0.07   & $85.4(104)$  \\ 

\chasa  & 4766 & 2.7 & $\le15.7$ & 1.70 &
$1.52^{+0.72}_{-0.46}$    &  45.52$\pm$0.16   &  $23.3(18)$  \\ 

\chasa  & 10306 {\sc and} 4766 & 2.7 & $\le4.8$  & 1.84$\pm$0.24 &
...        &  ... & $108.7(122)$  \\

\xmmsc & 3562 & 3.3 & $\le 3.4$ &  1.76$\pm$0.32 & 5.04$\pm$1.04 & 46.06$\pm$0.09 & 54.1(71) \\

\xmmsd & 3564 & 1.3 & $\le24.7$ & 1.70 & 2.43$\pm$0.71 & 45.71$\pm$0.13 & 24.0(26) \\

\chasb & 10305 & 1.4 & $\le0.6$ & 1.34$\pm$0.06 &
$12.78^{+0.57}_{-0.71}$ &  46.28$\pm$0.02   &  $443.9(508)$  \\ 

\chasb & 4769 & 1.4 & $\le3.2$ & 1.33$\pm$0.18 &
$7.15^{+1.01}_{-0.94}$    &   46.03$\pm$0.06  &  $154.2(159)$  \\ 

\chasb & 10305 {\sc and} 4769 & 1.4 &  $\le0.6$ & 1.34$\pm$0.06 &
...        &   ... & $598.1(668)$  \\

\enddata
\tablecomments{
The best-fit  column density,  were obtained from a model consisting of a Galactic-absorbed power-law with intrinsic absorption at the redshift of the source; the same model has been used to estimate the best-fit  photon index, and \Cstat\ for  the  observations of \xmmsa. 
The best-fit  photon index, and \Cstat\ of all the other observations were obtained from a model consisting of a Galactic absorbed \PL.
Errors and upper limits represent 90\% confidence intervals for each value,
taking one parameter to be of interest \citep[$\Delta C$=2.71; e.g.,][]{1976ApJ...210..642A}. }

\tablenotetext{a}{Galactic absorption column densities 
\citep[from][]{1990ARA&A..28..215D}, obtained at the optical coordinates of the
sources (see Tables~\ref{tab:xmmo} and \ref{tab:chao})  through the use of the HEASARC \nh\ tool at
http://heasarc.gsfc.nasa.gov/cgi-bin/Tools/w3nh/w3nh.pl}
\tablenotetext{b}{Intrinsic column density in units of 10$^{22}$~cm$^{-2}$.
}
\tablenotetext{c}{Absorption-corrected flux in the observed 0.5$-$2 keV band in units of $10^{-14}$~\flux. This flux is obtained from the best fitted parameters of an absorbed \PL\ fit. Note that the intrinsic absorption is only important for \xmmsa; for the \xmm\ observation of this source $f_{0.5-2}=(14.16\pm 0.75)\times 10^{-14}~\flux$ when we do not correct for intrinsic absorption.}

\tablenotetext{d}{\hbox{Logarithm} of the absorption-corrected  luminosity in the 2$-$10 keV rest-frame band; the luminosity is in units of \lumin. }

\end{deluxetable*}

We find a low-level secondary radial structure in
the 4.9~GHz image, which can be seen in the south western tip of the contour plots in the right panel of Figure~\ref{fig:char}. This structure is apparent more clearly when we super-resolve the image using a restoring beam of 0.6\arcsec$\times$0.6\arcsec\ (see the contour plot in Figure~\ref{fig:rama}).  The south western extension at 4.9~GHz has a peak flux density of 0.54~mJy per beam, around 5 times higher than the rms of 0.11~mJy per beam.  This 4.9~GHz extension in \chasb\ is spatially distinct
from the southernly extension claimed at 1.4 GHz by \cite{2006ApJ...650..679C,2008ASPC..386..462C}.
 At this southernly declination, the VLA synthesized beam at 1.4 GHz
has dimensions of 2.70\arcsec$\times$1.00\arcsec, with the major axis oriented north-south.
Cheung et al. use a restoring beam of 0.75\arcsec$\times$0.75\arcsec\ to deconvolve this
source, super-resolving the major axis by almost a factor of four. The
claimed secondary radio component extends to the south, in the direction
of maximum super-resolution and highest uncertainty. We have similarly
super-resolved the 1.4 GHz image of \chasb\ and are unable to
replicate the component shown by \cite{2006ApJ...650..679C,2008ASPC..386..462C}.
To compare agreement between the \XR\ extension and the secondary radio structure found at 4.9~GHz in \chasb, we  performed a \hbox{maximum-likelihood} reconstruction \citep[see \S2.1 of][]{2006AJ....131.2164T} on the full band image (250 iterations). 
The reconstructed image shows various marginal features (1--3\% of the
core intensity) at all position angles, although somewhat stronger to
the south and west. In Figure~\ref{fig:rama} we show a reconstructed image (yellow-red-blue indicate decreasing order of count density) with logarithmic intensity scaling. From Figure~\ref{fig:rama} we notice that the \XR\ extended emission of \chasb\ approximately matches the weak radio extension in the SW direction. 

\begin{figure*}
 \includegraphics[width=12.4cm]{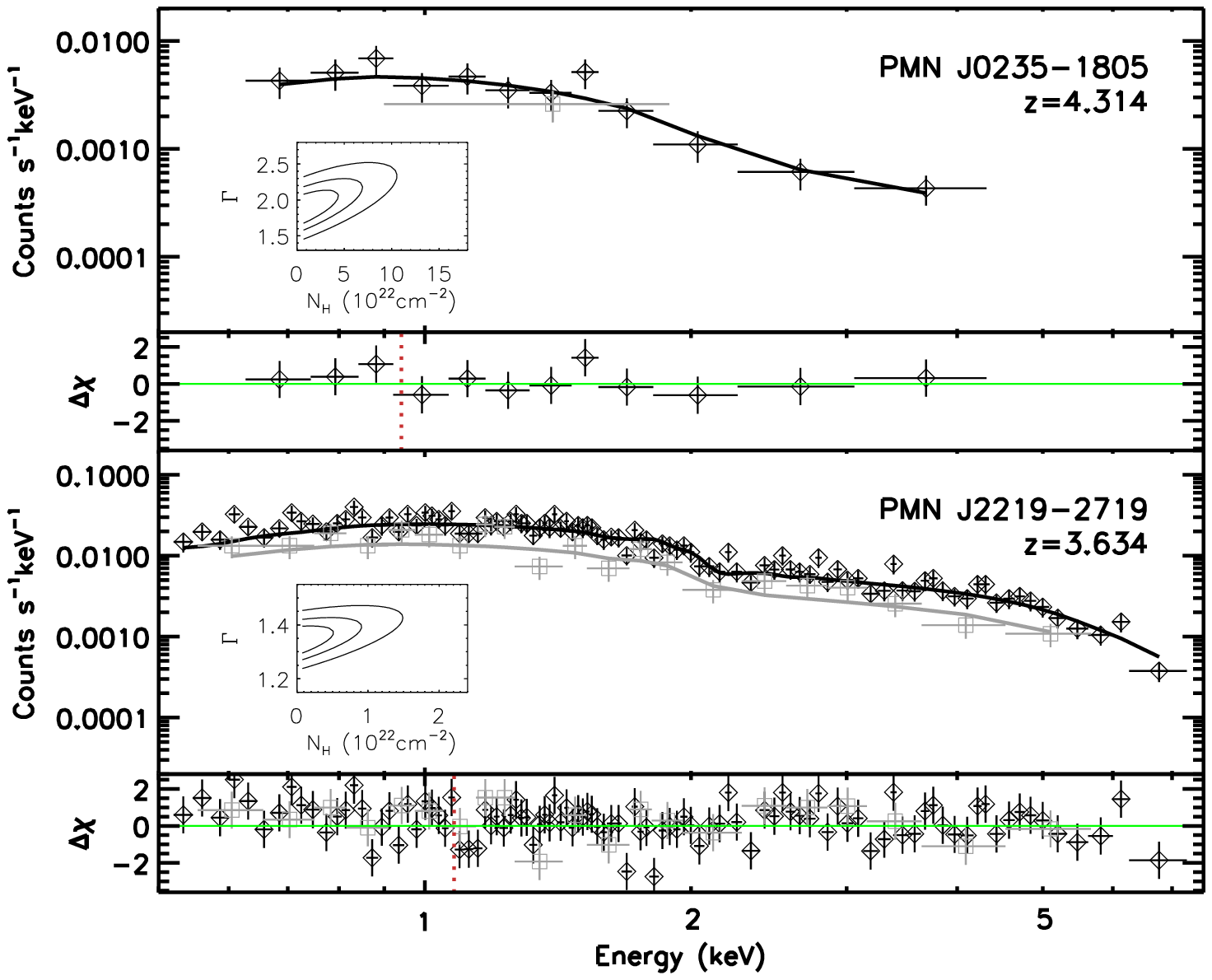}
        \centering
       \caption{
        \XR\ spectral fits of a power-law with Galactic absorption to the \chandra\ observations of \chasa\ and \chasb. The \hbox{best-fit} \PL\  to the data above 5~keV rest-frame  and extrapolated downward (solid line)  is shown. The open diamonds correspond to the data points of our most recent \chandra\ observations of  \chasa\ and \chasb.  The open squares correspond to  the data points of the \chandra\ observations of  \chasa\  and \chasb\  analyzed in \cite{2006AJ....131.1914L}. The $\Delta \chi$ residuals show the deviation of the data from the model in units of $\sigma$ with error bars of size unity. In each residual panel  we have marked with a dotted vertical line the 5~keV rest-frame energy.
        The inset in each panel shows 68\%, 90\%, and 99\% confidence contours for the intrinsic absorption ($N_{\rm H}$) and photon index ($\Gamma$) using the simultaneous \hbox{$C$-statistic} fits of observations 4766 and 10306  for \chasa\  and 4769 and 10305 for  \chasb.}
     \label{fig:chda}
     \end{figure*}
     
      \begin{figure*}
 \includegraphics[width=14cm,height=17cm]{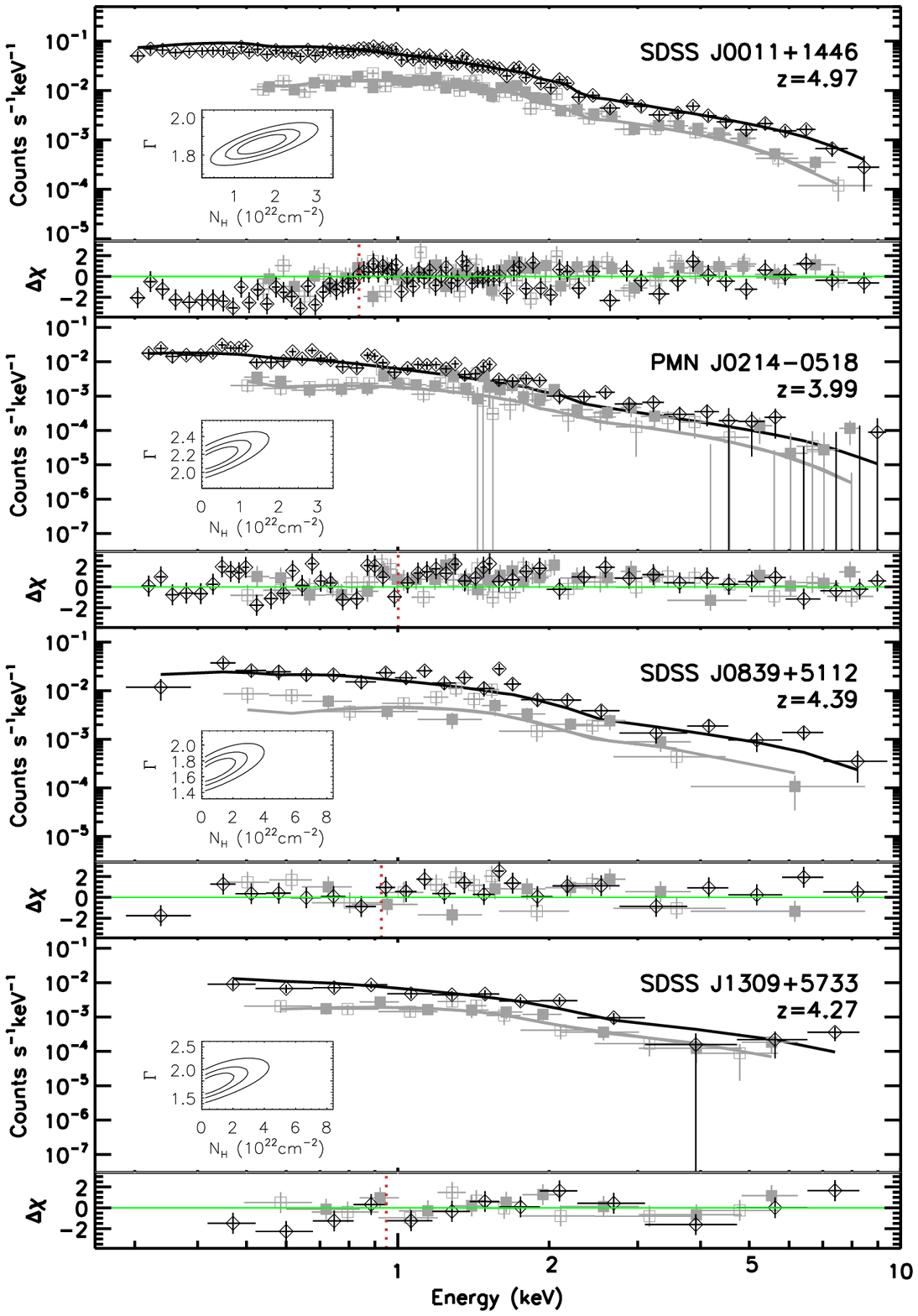}
        \centering
       \caption{ \XR\ spectral fits of a power-law with Galactic absorption to our new \xmm\ observations of $z \gtrsim 4$ RLQs. The \hbox{best-fit} \PL\  to the data above 5~keV rest-frame  and extrapolated downward in energy (solid line) is shown. The $\Delta \chi$ residuals show the deviation of the data from the model in units of $\sigma$ with error bars of size unity. The open diamonds, open squares, and filled squares correspond to the data points of the PN, MOS1 and MOS2 spectra, respectively, for our \xmm\ observations.
        In each residual panel  we have marked with a dotted vertical line the 5~keV rest-frame energy.  Note the systematic negative residuals  below $\approx 0.8$~keV for \xmmsa, corresponding to the detected intrinsic absorption.   The inset in each panel shows 68\%, 90\%, and 99\% confidence contours for the intrinsic absorption ($N_{\rm H}$) and photon index ($\Gamma$) using the simultaneous \Cstat\ fits of the pn, MOS1 and MOS2 data points.}
     \label{fig:xmda}
     \end{figure*}

Additionally, for the new \chandra\ observation of \chasa, the excess of \XR\ counts that lie at  $>$1\arcsec\ from the core were consistent with zero when compared with the \marx\ PSF.   The line-of-sight projected jet emission in \chasa\ and \chasb\ is $\lesssim 2.6\%$ of the core emission in these sources. Similarly, previous work on the $z = 4.30$ RLQ GB~1508$+$5714 demonstrated that the jet-to-core \XR\ flux ratio was $\sim$ 3\% \citep{2003ApJ...598L..15S, 2003MNRAS.346L...7Y}. The jet \XR\ emission in \chasb\ is also consistent with a mean of $\approx 4\%$ ($\sigma \approx 6 \%$) jet-to-core \XR\ flux ratio found  in  a sample (20 objects) of $z\lesssim2.6$ quasar jets  by \cite{2005ApJS..156...13M, 2011ApJS..193...15M}.
Therefore in neither of our new \chandra\ observations do we detect a fractionally large jet emission when compared to the core emission. As a consequence, we do not find any evidence of enhanced jet emission at the high redshifts of these sources as we might expect from the IC/CMB model \citep[see, e.g.,][]{2002ApJ...569L..23S}.

\subsection{Spectral analysis \label{S:Span}}

The spectral analysis was performed using \xspec\ v.12.0; the fitting bands used were  \hbox{0.5--8~keV} for the \chandra\ observations, \hbox{0.5--10~keV} for the \xmm\ MOS camera observations, and \hbox{0.3--10~keV} for the \xmm\ pn camera observations. 
For  \chandra\  we performed fits for each of the new and archival observations. Since the sources \chasa\ and \chasb\ have both new and archival observations with \chandra, we have performed a joint spectral fitting of each new observation with its respective archival observation. For these joint fits all the spectral parameters excluding the normalization are constrained to be the same. For the \xmm\ data we  performed fits on each of the EPIC camera spectra separately; we also performed joint fits using the three EPIC cameras. The fits for each EPIC camera separately are consistent with the joint fits, and therefore we present only the joint fits in this work.

One of the  objectives of the spectral analysis is to determine if intrinsic absorption is important in typical high-redshift RLQs. As a first approach  to assess the presence of absorption in our sources, we group the spectra to have between 10--30 counts per bin  in each observation/instrument. Using these grouped data we fit the observations (using  the $\chi^2$ statistic) above 5~keV in the rest frame (about 1 keV in the observed frame) with a \PL\  and then extrapolate it to lower energies. Through the use of this method, the residuals found at low energies may indicate absorption in the spectra. Evidence for absorption is only seen in the spectrum of \xmmsa\ (see Figures~\ref{fig:chda} and \ref{fig:xmda}). The qualitative analysis above is also consistent with our spectral fits.  These spectral fits were performed in the full fitting band of each instrument using the \Cstat\ \citep{1979ApJ...228..939C}.\footnote{No spectral grouping was used for the \Cstat\ fits.}
We first try a fit using a \PL\ model with Galactic absorption and intrinsic absorption (the \xspec\ model {\sc phabs*zphabs*pow}). The Galactic absorption column densities were obtained from \cite{1990ARA&A..28..215D};  the errors on the Galactic column densities range between $(1-3)\times 10^{19}$~\cmsq\ \citep{1994ApJS...95..413E, 1996ApJS..105..369M}.   In all the fits solar abundances were assumed for the Galactic and intrinsic absorption.  We do not find  evidence of significant intrinsic absorption in five out of our six new RLQ observations. This can be inferred from the best-fitted parameters obtained from the absorbed \PL\ model. The column densities for these five sources are  consistent with zero. 
Additionally, contour plots of $\Gamma$ vs. \nh\  indicate that a null intrinsic column density is inside the 68\% confidence region (see Figures~\ref{fig:chda} and \ref{fig:xmda}). 
The source that shows evidence for absorption at $>$99\% confidence is \xmmsa.  The absorption found in this source is of moderate strength ($\nh \approx 1.7 \times 10^{22}~\cmsq)$, in agreement with the absorption found in other high-redshift RLQs  \citep[e.g.,][]{2005MNRAS.364..195P, 2006MNRAS.368..985Y}.  To verify that the absorption detected cannot be reproduced by unaccounted uncertainties in the Galactic absorption, we have fitted the \xmm\ spectra (joint fit of the EPIC cameras) of  \xmmsa\ using a Galactic absorbed \PL\ model  (i.e., the \xspec\ model {\sc phabs*pow}).  The best-fitted  absorption obtained, with the absorption at $z=0$ as a free parameter, is $(7.7\pm1.5)\times10^{20}$~\cmsq\ (error at the 90\% confidence level), well above the expected value for Galactic absorption (see Table~\ref{tab:xfit}).   Based on the fact that just one observation presents absorption, we re-fit all the other observations (new and archival) with a \PL\ model with only Galactic absorption (the \xspec\ model {\sc phabs*pow}). The best-fit spectral parameters (errors and upper limits at the 90\% confidence level) are presented in Table~\ref{tab:xfit}. In this table, the best-fit photon index and \Cstat, of all the observations that do not present absorption, are obtained from a model consisting of a Galactic-absorbed \PL.  In the case of  \xmmsa\   the best-fit photon index and \Cstat\ presented in Table~\ref{tab:xfit} are obtained from a model consisting of a Galactic-absorbed \PL\ with intrinsic  absorption at the redshift of the source. 
 The mean photon index for our sample is $1.74\pm0.11$ (1$\sigma$ error on the mean), in agreement with samples of \hbox{low-redshift} RLQs \citep{1999ApJ...526...60S, 2000MNRAS.316..234R}. Based on this, in Table~\ref{tab:xfit} for the \chandra\ archival observations of \chasa\ and \xmmsd\  we have fixed the \PL\ photon index to $\Gamma=1.7$ in order to avoid degeneracies associated with the limited numbers of counts ($\lesssim 30$; see Table~\ref{tab:chao}) in these observations.

\begin{deluxetable*}{ccccccc}
\tabletypesize{\scriptsize}
\tablecolumns{5} \tablewidth{0pt} \tablecaption{Two epoch X-ray flux variability test.
\label{tab:varb}}

\tablehead{
\colhead{} & \colhead{$f_{0.5-2}$\tablenotemark{a}} &
\colhead{$f_{0.5-2}$\tablenotemark{a}} & \colhead{$\Delta t$\tablenotemark{b}}  \\
\colhead{\sc object name} & \colhead{(epoch 1)}  &
\colhead{(epoch 2)} & \colhead{(days)} & \colhead{$\chi^2$}  & 
 \colhead{\%sig\tablenotemark{c}}
}

\startdata

\xmmsa & $12.31^{+2.04}_{-1.09}$ & 13.74$\pm$0.42 & 406 & 0.5 & 
 51.1\\
\xmmsb & $3.72^{+0.77}_{-0.52}$ & 1.88$\pm 0.11$  & 446  &11.98  & 
$>$99.9\\
\chasa &   $1.52^{+0.46}_{-0.27}$ & 2.32$\pm$0.24 & 347 & 2.4 & 
87.8 \\

\xmmsc & 5.04$\pm${-0.64} &  $4.42^{+0.47}_{-0.32}$ & 218 & 0.6  & 56.8 \\

\xmmsd & $2.43^{+0.39}_{-0.47}$ &  1.68$\pm$0.16 & 193 & 2.3 &   87.0 \\
\chasb &    7.15$\pm$0.59 & $12.78^{+0.32}_{-0.48}$ & 459 & 54.8 & 
$>$99.9\\

\enddata


\tablenotetext{a}{Absorption-corrected flux in the observed 0.5$-$2 keV band in units of $10^{-14}$~\flux. 
For each source the fluxes at epoch~1 and epoch~2 correspond to those obtained from the  \chandra\ archival observations and from the new observations (either with \chandra\ or \xmm), respectively. }

\tablenotetext{b}{Rest-frame timescale.}


\tablenotetext{c}{Percentile of significance.}

\end{deluxetable*}

  Since the average redshift of our sources is close to four, the \RF\ band of the fitting  for each source reaches  energies up to $\sim 50$~keV. Therefore, the source spectra are sampling any potential \hbox{Compton-reflection} component which peaks at \RF\ energies of $\sim 30$~keV \citep[e.g.,][]{1988ApJ...335...57L, 1995MNRAS.273..837M}. In general we would not expect reflection signatures to be strong in these sources due to likely dilution by the jet-linked \XR\ continuum. To check for reflection we fitted our spectra above 5~keV in the \RF\ using the {\sc pexrav} model. From these fits, we find that the reflection parameter is consistent with zero in all our observations, and therefore we do not find that reflection is significant in any of our sources.  We also used Gaussian line profiles to fit our spectra in the energy range where the Fe~K$\alpha$ line is found (6.4~keV \RF\ for neutral iron). Through this procedure we   find that the Fe K$\alpha$ best-fitted equivalent widths (EWs) are consistent with zero for all our observations, and therefore there is no evidence of Fe K$\alpha$ signatures in our spectra.  For the new observation of \xmmsa, which is the observation of highest S/N, we obtain that the EW of Fe~K$\alpha$ is $\lesssim 170$~eV. The upper limit on the EW of Fe~K$\alpha$ line for the other sources is less well constrained. The absence of  significant Fe~K$\alpha$ emission lines is consistent with what is expected for luminous ($\lx \gtrsim 10^{46}~\lumin$; Table~\ref{tab:xfit}) quasars  \citep[e.g.,][]{1997ApJ...488L..91N, 2004MNRAS.347..316P,2007A&A...467L..19B}. 
 
 \begin{figure*}
   \includegraphics[width=16cm]{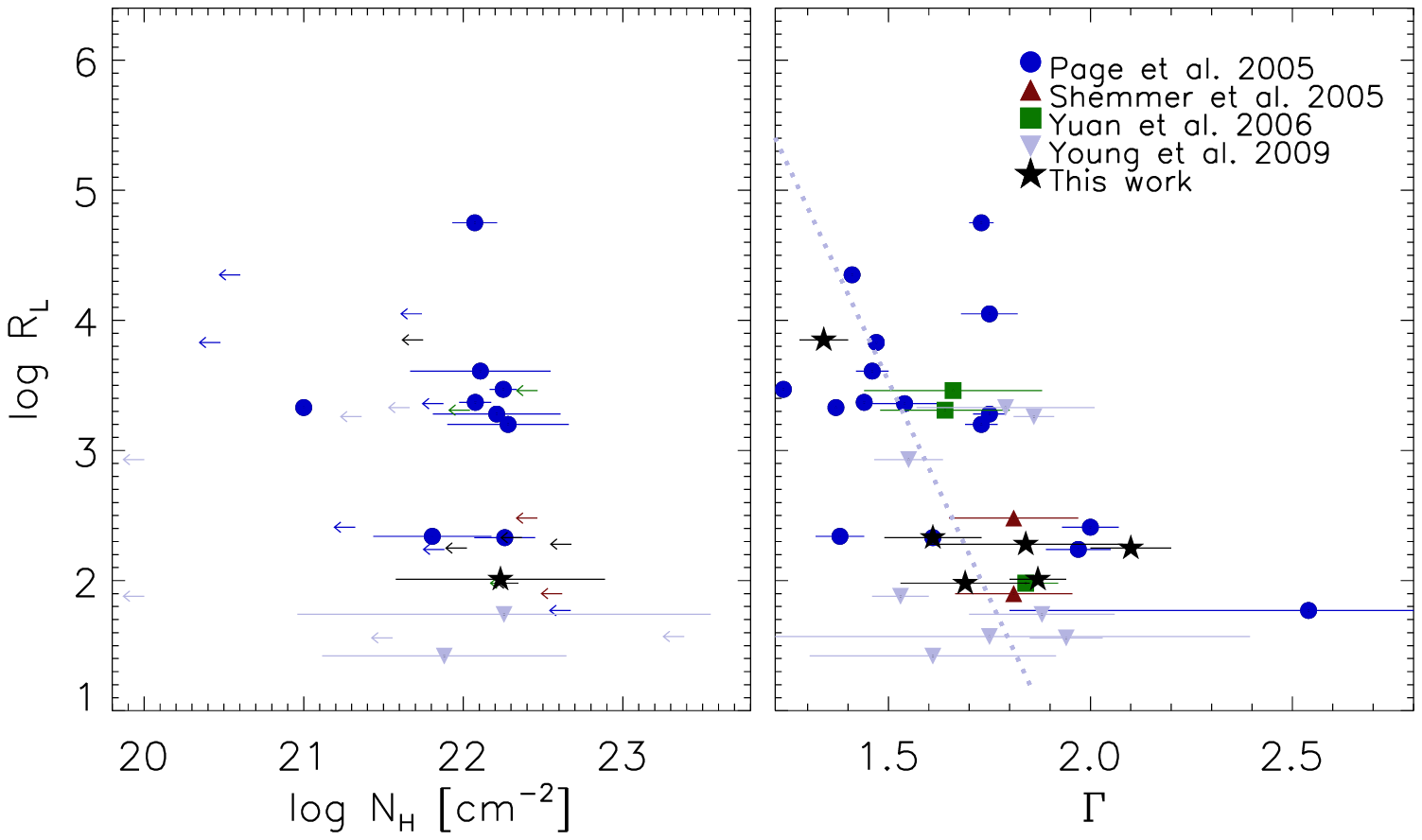}
        \centering
       \caption{Radio-loudness parameter versus best fitted column density ($N_{\rm H}$; left panel) and photon index ($\Gamma$; right panel)  for our sample (stars) and $z>2.0$ RLQ observations with \chandra\ and \xmm\ (circles are from \citeauthor{2005MNRAS.364..195P}~2005, triangles from \citeauthor{2005ApJ...630..729S}~2005, squares from \citeauthor{2006MNRAS.368..985Y}~2006, and inverted triangles from  \citeauthor{2009ApJS..183...17Y}~2009).  
 The arrows in the left panel correspond to column density upper limits. 
 The dotted line is obtained by using the the  IDL Astronomy library tool {\sc linmix\_err} \citep{2007ApJ...665.1489K} on our enlarged sample.}
     \label{fig:RLsa}
     \end{figure*}

\subsection{Short and long-term variability  \label{S:vari}}

We searched for rapid variability in the new \xmm\ and \chandra\  observations by applying the Kolmogorov-Smirnov test to the photon arrival times; however, no significant variability was found.
Long-term variability was assessed by comparing the fluxes in the \hbox{0.5--2~keV} band from the archival observations with those of the new observations. Comparing the fluxes in the two observation epochs of each source (Table~\ref{tab:xfit}), we find that one third of our sample (2 out of 6) shows potential variability. The sources which possibly present variability are \xmmsb, and \chasb. In order to test the significance of this variability, we calculate for each source the $\chi^2$ statistic where the data points are the fluxes of each observation (with their respective 1$\sigma$ errors) and the model is a constant flux with the best fit of the two epochs. The $\chi^2$ value provides a statistical test of the null hypothesis that the flux of each epoch is equal to the best-fitted flux of the two epochs. In Table~\ref{tab:varb} we list the fluxes for each source in the \chandra\ archival observations (epoch~1) and in the new observations (epoch~2). In this table we also have listed the values of  $\chi^2$ for our model to test variability. 
  For \xmmsb,  and \chasb\ we reject the null hypothesis with $>99$\% significance.\footnote{We arrive to the same conclusion if we compare the  fluxes in the \hbox{0.5--8~keV} band from the archival observations with those of the new observations.}   The \RF\ time scales for the variability  are $446$~days and $459$~days for \xmmsb\  and \chasb, respectively. The amplitude factors of the flux variability are  $\approx$2.0 and $\approx$1.8 for \xmmsb\  and \chasb, respectively; these are in agreement with the typical amplitude factors of lower redshift RLQs \citep[e.g.,][]{1997A&A...319..413B}. The two sources that show evidence  for variability are the two with the longest time scales between observations, but this may be merely coincidental.

\subsection{Power-law and absorption correlation tests \label{S:corr}}

In order to investigate possible relations between \XR\ spectral parameters, redshift, and $R_L$, we have enlarged our original sample using some of the latest  high-redshift ($z>2$) RLQ \XR\ measurements.
The $R_L$ vs. $z$ diagram of the enlarged sample can be seen in Figure~\ref{fig:zRLs} in which the sources from each survey have been indicated with different symbols. 
We have also plotted in Figure~\ref{fig:RLsa}  diagrams of $R_L$ vs $N_H$ (left panel) and $R_L$ vs. $\Gamma$ (right panel) for RLQs at $z>2$.  From the right panel we infer a likely decrease of $\Gamma$ with increasing $R_L$.  To quantify possible dependences between the parameters, we have calculated Spearman correlation coefficients of $\Gamma$ vs. $R_L$,  $\Gamma$ vs. $z$, \nh\ vs. $R_L$, and also \nh\ vs. $z$; the results are given in Table~\ref{tab:corr} considering sources with $z>2$ and $z>3$ separately. We find an anticorrelation between $\Gamma$ and $R_L$; this anticorrelation is significant (at the $99.9$\% confidence level) for RLQs with $z > 2$ and with $z > 3$.  The hardening of the spectra with increasing $R_L$ indicates a rising jet contribution to the \XR\ spectra as a function of $R_L$. This conclusion is reinforced by the fact that the excess \XR\ luminosity  in RLQs (with respect to RQQs) increases with $R_L$ as quantified by, e.g., \cite{2011ApJ...726...20M}.
Notice that, with  the exception of  \chasb,  our sources possess moderate values of $R_L$ (see Table~\ref{tab:opra}). The photon index for \chasb\ of $\Gamma \approx 1.3$ is clearly  harder when compared with the photon indices of the other RLQs in our sample (on average $\approx 1.7$; see \S \ref{S:Span}); this is in agreement with the trend found. Additionally,  the soft spectral index  ($\Gamma \approx 2.1$) of  \xmmsb\ ($\lRL\sim2$) is also  in accordance with the trend found (see Figure~\ref{fig:RLsa}).  

The Pearson correlation coefficient of $\Gamma$ with \lRL\  for our extended sample is $-$0.487 (significant at the 99.7 \% confidence level); therefore, as a second step we fit the relation between $\Gamma$ and \lRL\ with a simple linear model. This fit is not statistically acceptable because the error bars on $\Gamma$ are not representative of the intrinsic dispersion of the data around a possible  $\Gamma-\lRL$ relation.  In view of this, we adopt the  IDL Astronomy Library tool {\sc linmix\_err}  \citep{2007ApJ...665.1489K} to estimate the parameters describing a linear model between  $\Gamma$ and \lRL\  (i.e.,  find $\alpha$ and $\beta$ of $\Gamma=\alpha+\beta \times \lRL$). This IDL routine, which accounts for measurement errors, nondetections, and intrinsic scatter,  is implemented with a Bayesian approach to compute the posterior probability distribution from the observed data.  From the random draws ($2\times10^5$ realizations) from the posterior probability distribution we find that the median  values of the parameters $\alpha$ and $\beta$ are given by

\begin{equation} \label{eq:lfit}
 \Gamma= (2.03 \pm 0.11)+(-0.15 \pm 0.04)\times \lRL. 
\end{equation}

    \begin{figure}
   \includegraphics[width=8cm]{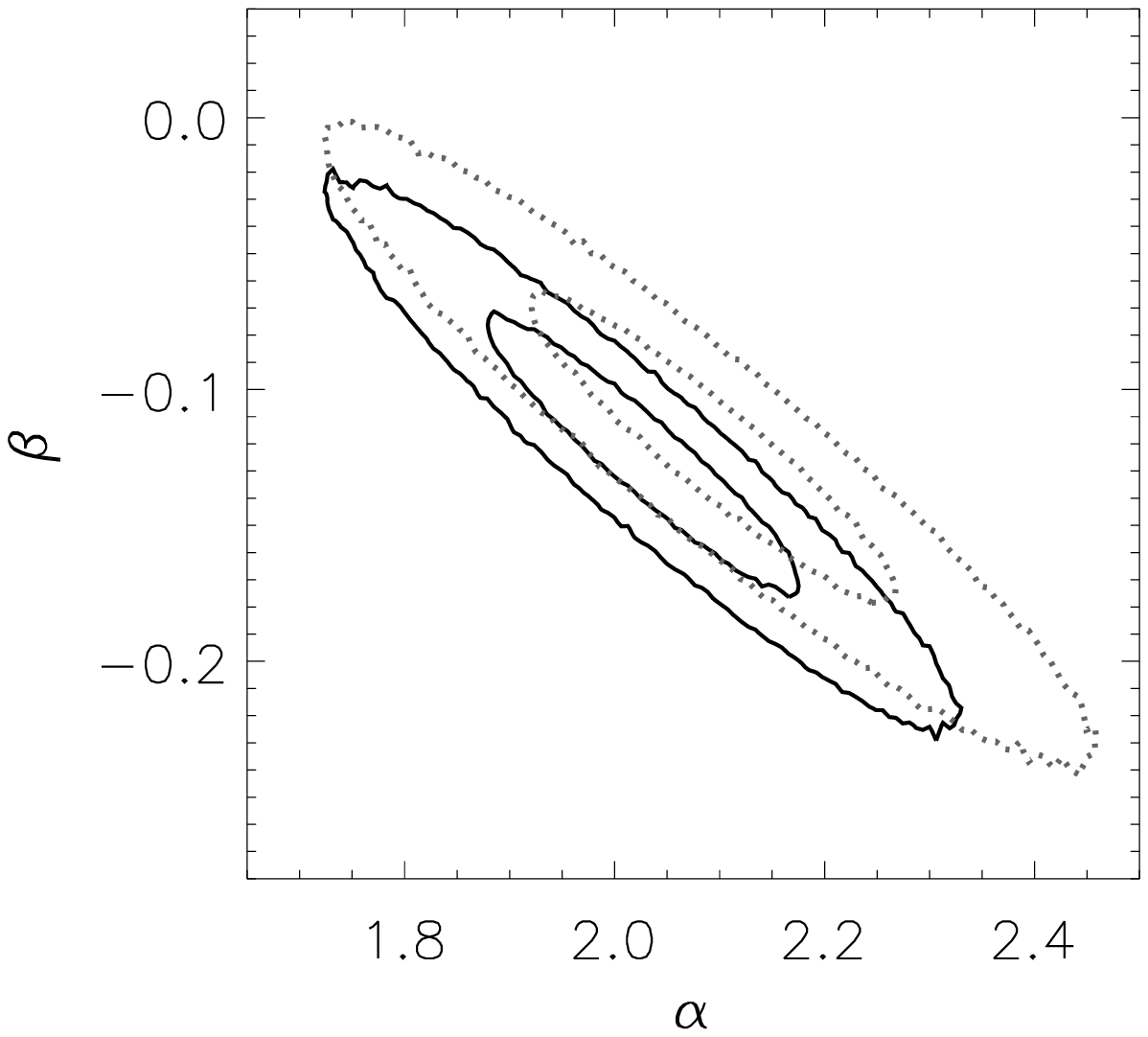}
        \centering
         \caption{Contour plots of $\alpha$ vs. $\beta$ for the model $\Gamma=\alpha+\beta  \times  \lRL$  using the the  IDL Astronomy Library tool {\sc linmix\_err}  \citep{2007ApJ...665.1489K}. The solid and dotted contours are the 68\% and 99\% confidence regions of  $2\times10^5$ realizations for our and the \citeauthor{2000MNRAS.316..234R} data, respectively.}
        
     \label{fig:Reev}
     \end{figure}

The \hbox{$\Gamma-\lRL$} relation has already been found in the past in $z \lesssim 2$ samples of RLQs \citep[e.g.,][]{1987ApJ...323..243W, 2000MNRAS.316..234R}. To test if the relation found in equation~\ref{eq:lfit} is consistent with that found at lower redshifts, we  selected $z<2$ RLQs from \cite{2000MNRAS.316..234R}. Using  {\sc linmix\_err} ($2\times10^5$ realizations)  we find that the median  values of the parameters $\alpha$ and $\beta$ are given by
\begin{equation}  \label{eq:lfRe}
\Gamma= (2.09 \pm 0.16)+(-0.12 \pm 0.05)\times \lRL.
\end{equation}

The linear parameters obtained for the $z<2$ sources of  \citeauthor{2000MNRAS.316..234R}  (equation~\ref{eq:lfRe}) are consistent with those found for our sources (see equation 1, and Figure~\ref{fig:Reev}); therefore,  we do not find evidence that  $\Gamma-\lRL$ relation changes with redshift.

The correlation tests in Table~\ref{tab:corr} give indications that $N_H$ could be increasing with $z$ for sources with $z > 2$ \citep[in agreement with][]{2000MNRAS.316..234R, 2005MNRAS.364..195P}. The correlation of \nh\ with $z$ is  at the $\sim 97$\% confidence level.  
Therefore, an apparent correlation between $\nh$ and $z$ may be the result of observational bias, since the threshold of absorption that can be measured increases with redshift.\footnote{  The minimum rest-frame energy of the observed X-ray band increases as (1+z), e.g., for $z\sim4$ the minimum \RF\ energy of the observed band is $\gtrsim 2$~keV. A typical intrinsic absorption of column density \hbox{$\sim10^{22}$~\cmsq} shows a signature in the spectra at \RF\ energies $\lesssim 2$~keV. Therefore,  the threshold absorption that will produce a significant feature in the spectrum increases with redshift. } 

\begin{deluxetable}{ccccccc}
\tabletypesize{\scriptsize}
\tablecolumns{5} \tablewidth{0pt} \tablecaption{Correlation analysis results for a combined sample of high-redshift radio-loud quasars.
\label{tab:corr}}

\tablehead{
\colhead{Tested parameters} & \colhead{Redshift range} &
\colhead{$N$\tablenotemark{a}} & \colhead{$r_S$\tablenotemark{b}} &
\colhead{\% sig\tablenotemark{c}} 
}

\startdata

$\Gamma$~vs.~$R_L$ & $z > 2.0$ & 35 & $-0.542$ & 99.9 \\
$\Gamma$~vs.~$R_L$ & $z > 3.0$ & 22 & $-0.637$ & 99.9 \\
$\Gamma$~vs.~$z$ & $z > 2.0$ & 35 & $0.063$ & 28.1 \\
$\Gamma$~vs.~$z$ & $z > 3.0$ & 22 & $0.341$ & 88.0 \\
\nh~vs.~$R_L$\tablenotemark{d} & $z > 2.0$ & 35 & $-0.165$ & 66.4 \\
\nh~vs.~$R_L$\tablenotemark{d} & $z > 3.0$ & 22 & $-0.344$  & 88.5 \\
\nh~vs.~$z$\tablenotemark{d} & $z > 2.0$ & 35 & 0.370 & 97.0\\
\nh~vs.~$z$\tablenotemark{d} & $z > 3.0$ &  22 & 0.395 & 93.0 \\

\enddata

\tablecomments{The sample used to calculate the correlation parameters corresponds to radio-loud quasars presented in Tables~\ref{tab:xmmo} and \ref{tab:chao},  and those extracted from \cite{2005MNRAS.364..195P}, \cite{2005ApJ...630..729S}, \cite{2006MNRAS.368..985Y}, and   \cite{2009ApJS..183...17Y}. }

\tablenotetext{a}{Number of RL quasars in each redshift bin. }

\tablenotetext{b}{Spearman correlation coefficient. }

\tablenotetext{c}{Percentile significance of the correlation.}

\tablenotetext{d}{The correlation coefficients involving \nh\ were calculated using survival analysis with the ASURV code \citep{1986ApJ...306..490I}. \\ }

\end{deluxetable}

\section{SUMMARY AND CONCLUSIONS}

We have reported the \XR\ properties of six high-redshift ($z \gtrsim 4$) RLQs,
observed with \xmm\ (four observations) and \chandra\ (two observations) as a 
follow-up of earlier \chandra\ snapshot observations. A key aspect of our 
observed sample is that it spans lower values of radio loudness than past 
studies of high-redshift RLQs in the \XR\ regime; the moderate radio-loudness
objects studied are much more representative of the overall RLQ population.  
The main conclusions from our spectral, imaging, and variability analyses are 
the following: 

\begin{enumerate}

\item 
The power-law \XR\ continua of our targets have a mean photon index of 
$\langle \Gamma \rangle=1.74\pm0.11$, consistent with measurements of lower redshift RLQs. 
They also follow an anti-correlation between photon index and radio loudness
seen at lower redshifts. Using linear fits to the $\Gamma-\lRL$ relation for an extended sample at $z>2$, we find consistency of this dependence in relation to lower redshift ($z<2$) samples.
We find no evidence of Compton-reflection continua 
or iron~K$\alpha$ emission lines. The measured continuum shapes, combined with the 
fact that our targets are on average $\approx 6$ times brighter in \hbox{X-rays} 
than RQQs of similar UV luminosity, indicate that jet-linked emission 
dominates the \XR\ continua.  

\item 
We find evidence that one of our targets, \xmmsa, has significant \XR\
absorption with a column density of $N_{\rm H}\approx 1.7\times 10^{22}~\cmsq$. 
We set useful limits upon \XR\ absorption for the rest. The incidence of 
\XR\ absorption in our sample appears roughly consistent with that for 
more highly radio-loud objects at similar redshifts. 

\item 
In the \chandra\ observation of \chasb\ we detect a likely \XR\
jet with an extension of $\sim 14$~kpc. The \XR\ luminosity of this 
putative jet is $\approx 2$\% that of the core. 
In the \chandra\ observations of \chasa\ and \chasb\ we do not find evidence of  enhanced  extended \XR\ jet emission as we might expect  from the IC/CMB model. We do not find any evidence of radio extension in the 1.4 and 4.9 GHz 
VLA images of \chasa.  We find a secondary structure in the 4.9~GHz VLA image of \chasb\ that matches the \XR\ extension found in this source; due to low resolution we are unable to see this structure in the 1.4~GHz VLA image.

\item 
Two of our targets, PMN~J0214--0518 and PMN~J2219--2719, show evidence
for \XR\ variability by \hbox{80--100\%} on rest-frame timescales of 
\hbox{450--460~days}. 

\end{enumerate}

\acknowledgments

We thank M. Young, B. Kelly, C. C. Cheung and the anonymous referee for helpful discussions and comments. 
We acknowledge support from NASA ADP grant NNX10AC99G (CS, WNB), Chandra \XR\ Center grant \hbox{G09-0112X} (CS, WNB), and an AAUW
American Dissertation Fellowship (LAL). C.~Vignali acknowledges support from the Italian Space Agency (ASI) under the \hbox{ASI-INAF} contracts I/009/10/0 and I/088/06/0.  B.~P.~Miller acknowledges support by \chandra\ Award Number GO0-11112A. The National Radio Astronomy Observatory is a facility of the National Science Foundation operated under cooperative agreement by Associated Universities, Inc.

\bibliographystyle{apj}
\bibliography{msbib}

\end{document}